\documentclass{iopart}
\usepackage{amssymb,mathdots}
\usepackage{iopams}
\usepackage{amsthm}
\usepackage{amscd}
\usepackage{amsfonts}
\usepackage{dsfont}
\usepackage{cite}
\usepackage{graphicx}

\def\be{\begin{equation}}
\def\ee{\end{equation}}
\def\ba{\begin{eqnarray}}
\def\ea{\end{eqnarray}}

\def\ra{\rightarrow}

\def\ni{\noindent}
\def\pt{$\mathcal{PT}$}
\def\cpt{C$\mathcal{PT}$}

\def\){\right)}
\def\({\left(}

\begin{document}
\title{Metric operators for non-Hermitian quadratic su(2) Hamiltonians}
\author{Paulo E. G. Assis}
\address{
School of Mathematics, Statistics and Actuarial Sciences, \\
University of Kent, Canterbury, CT2 7NZ, UK}
\eads{\mailto{peg@kent.ac.uk}}
\begin{abstract}
A class of non-Hermitian quadratic $su(2)$ Hamiltonians having an anti-linear symmetry is constructed.
This is achieved by analysing the possible symmetries of such systems in terms of automorphisms of the algebra.
In fact, different realisations for this type of symmetry are obtained, including the natural occurrence of charge conjugation together with  parity and time reversal. Once specified the underlying anti-linear symmetry of the Hamiltonian, the former, if unbroken, leads to a purely real spectrum and the latter can be mapped to a Hermitian counterpart by, amongst other possibilities, a similarity transformation. Here, Lie-algebraic methods which were used to investigate the generalised Swanson Hamiltonian \cite{afkk} are employed to identify the class of quadratic Hamiltonians that allow for such a mapping to the Hermitian counterpart. Whereas for the linear $su(2)$ system every Hamiltonian of this type can be mapped to a Hermitian counterpart by a transformation which is itself an exponential of a linear combination of $su(2)$ generators, the situation is more complicated for quadratic Hamiltonians. Therefore, the possibility of more elaborate similarity transformations, including quadratic exponents, is also explored in detail. The existence of finite dimensional representations for the $su(2)$ Hamiltonian, as opposed to the $su(1,1)$ studied before, allows for comparison with explicit diagonalisation results for finite matrices. Finally, the similarity transformations constructed are compared with the analogue of Swanson's method for exact diagonalsation of the problem, establishing a simple relation between both approaches.
\end{abstract}\vspace{-3ex}
\qquad\qquad\qquad  J. Phys. A: Math. Theor. 44 265303 

\section{Introduction}

The interest in non-Hermitian Hamiltonians dates back to the early days of quantum mechanics. Traditionally these operators are associated with open systems, but, under additional conditions they can also be used to describe non-dissipative phenomena, e.g. \cite{assisbook, moiseyev}. Examples of such models have appeared in many areas, for instance in affine Toda theories \cite{hollowood, olive}, Regge field theory \cite{regge}, atomic physics \cite{scholtzgeyerhahne}, and also quantum spin chains \cite{von}. In the mathematical community, non-Hermitian operators had already been studied \cite{dieudonne, williamsadjoint, caliceti}, but the more recent work of Bender and Boettcher \cite{benderboettcher}, connecting the reality of quantum spectra with an unbroken symmetry under combined parity and time reversal, led to a wide-spread interest in so-called \pt-symmetric quantum systems. Subsequently, it has been shown \cite{Bend02b} that unbroken \pt-symmetry allows to introduce a new inner product with respect to which the time evolution generated by the non-Hermitian  \pt-symmetric Hamiltonian is unitary. The theoretical interest in non-Hermitian \pt-symmetric system also extends to classical physics, e.g. \cite{mostcomppot, benderholmhook, afcalogero, kaushal}.
Beyond mathematical physics motivations, recent activity in this area including experimental realisations of \pt-symmetric waveguides \cite{ptwave} boosted the interest in the field. Possible applications are now envisaged ranging from optics \cite{ptoptlat,optlat,optstruc} to condensed matter systems \cite{ptcondmat} and cold atoms \cite{ptbec,ptkahler}.

The key feature of \pt-symmetric quantum mechanics is that real spectra and unitary evolution can be obtained for Hamiltonians not satisfying the Hermiticity condition. The connection between an unbroken \pt-symmetry and a real spectrum is a special case of a more general theorem for non-Hermitian Hamiltonians presenting any anti-linear symmetry \cite{wigner, bberry}. If a Hamiltonian system possesses an anti-linear symmetry $A$,
\begin{equation}\label{sym}
[H, A] =0,
\end{equation}

\noindent and its eigenvectors are also invariant under this symmetry,
\begin{equation}\label{symbr}
A| \zeta_n \rangle = | \zeta_n \rangle,
\end{equation}

\noindent we say that the symmetry is in its unbroken regime. Then, because $\langle \xi | A^\dag A | \psi \rangle = \langle \psi  | \xi \rangle$, one can easily prove the reality of its spectrum without any reference to its Hermiticity properties \cite{wigner}, \cite{benderboettcher}:
\begin{equation}\label{realspec}
E_n | \zeta_n \rangle = H | \zeta_n \rangle = H A | \zeta_n \rangle = A H | \zeta_n \rangle = A \left( E_n | \zeta_n \rangle \right) = E_n^\ast | \zeta_n \rangle.
\end{equation}

If a system satisfies (\ref{sym}) but not (\ref{symbr}), one says it is in a phase of broken symmetry. In this regime eigenvalues are not necessarily real anymore but, instead, come in complex conjugate pairs. \footnote{Note, however, that although the coalescence of the complex conjugate pairs into real values cannot be guaranteed, the latter might still occur, eventually.} A widespread realisation in the community is to consider the anti-linear symmetry as the $\mathcal{PT}$ operator symbolizing simultaneous parity and time reversal transformations, together with complex conjugation.

Nonetheless, having the reality of the spectrum established is not enough for a quantum theory. One still needs to guarantee its unitary evolution. If the system is non-Hermitian, $H^\dag \neq H$ and a new Hermitian operator $\rho$, with respect to which the operator is Hermitian \cite{scholtzgeyerhahne},
\begin{equation}\label{genherm}
H^\dag \rho = \rho \, H,
\end{equation}
must be included. This operator has the effect of changing the scalar product in the underlying vector space. Besides, it allows the construction of the $\eta$ operator as $\rho = \eta^\dag \, \eta$, called the Dyson map, which makes possible the interpretation of the operators in the new Hilbert space, $X$ and $P$, for instance, rather than $x$ and $p$, as the genuine observables,
\begin{equation}
H(x,p) = \eta^{-1} h(x,p) \, \eta=  h(\eta^{-1}x\,\eta,\eta^{-1}p \,\eta) = h(X,P),
\end{equation}
\noindent with the fundamental commutation relation $[x,p] = [X,P] = {i} \hbar$ preserved. In this case the non-Hermitian Hamiltonian $ H$ can be mapped to an isospectral Hermitian Hamiltonian via a similarity transformation
\begin{equation}\label{dysonherm}
h = \eta H \eta^{-1} = h^\dag.
\end{equation}

According to the standard terminology, the Hamiltonian is said to be quasi-Hermitian if there is a Hermitian and positive metric operator $\rho$ satisfying (\ref{genherm}), whereas for $H$ to be considered pseudo-Hermitian $\rho$ must be Hermitian and invertible \cite{scholtzgeyerhahne, mostafazadeh}. The problem with the latter definition is that the definiteness of the metric is not guaranteed but with the former definition alone one cannot make a conclusive statement about the reality of the spectrum of the Hamiltonian.  For more details about \pt-symmetry and pseudo- and quasi-Hermitian Hamiltonians see e.g. \cite{sense, mostarev, mostarep, assisbook}. To circumvent these problems we impose simultaneously both requirements, leading to what will be henceforth referred to as pseudo-quasi-Hermiticity.

Assuming that the $\eta$ operator has an exponential form, $\eta = e^T$ for some operator $T$, the metric is automatically invertible and positive definite so that the reality of the spectra is guaranteed.
The Dyson operator $\eta$, therefore, constitutes a central element in the analysis of non-Hermitian systems. However, it is not unique and neither is the metric with respect to which the system is Hermitian according to (\ref{genherm}). As a trivial example, one can multiply the metric by any unitary operator, as this leaves the Hermiticity of the Hamiltonian unaffected. Similarly, multiplying it by any operator that commutes with the Hamiltonian would yield a new metric. Nevertheless, we shall not be concerned with ambiguity issues for the moment. Instead, we will focus on a systematic method to construct the necessary elements for a fundamental description of the system, which is not trivial. There is in fact only a handful of known models for which the analytical non-approximative construction of an $\eta$, or a metric, is possible. Often perturbative methods can be used to construct a metric for a non-Hermitian operator, e.g. \cite{caliperturb}, but other approaches have been used, notably in \cite{swanson, scholtzgeyer2, bagchiquesnesusy, ffmoyal, afmoyal, bila, musumbu, quesne, afkk, olalla, korffweston, siegl}.

The vast majority of investigations on \pt-symmerty and non-Hermitian operators focusses on one-dimensional quantum systems whose Hamiltonians are composed of coordinate and momentum operators, that is, operators from a Heisenberg-Weyl algebra. There are some investigations on algebraic methods for $su(1,1)$ \cite{quesne,afkk} and recently also the $E2$ algebra  \cite{BenderKalvkes}. In Hermitian quantum mechanics an important class of Hamiltonians is composed of $su(2)$ operators. These Hamiltonians have applications spanning diverse fields, reaching from the description of nuclear spins \cite{Lipkin} to ultra-cold atoms \cite{becsu2}. The spectral features of a \pt-symmetric generalisation of a special example have been studied in \cite{ptbec}.

In the present paper we investigate a general class of non-Hermitian Hamiltonians that are quadratic in the $su(2)$ operators, and construct the corresponding $\eta$ operators, in a continuation of work involving $su(1,1)$ quadratic Hamiltonians started in \cite{afkk}.
In the following section we report on the symmetries of the $su(2)$ generators and present the possible anti-linear symmetries underlying $su(2)$ Hamiltonians.  As a by-product we can show in a unified manner that both $su(1,1)$ and $su(2)$ algebras allow for three different symmetries, associated to the automorphisms of the algebra.  We demonstrate that certain systems of interest present, for instance,  a \cpt-symmetry, yet to be defined, rather than just \pt-symmetry.
In the sequence, in section \ref{secmet}, we tackle the problem of finding $\eta$ for the quadratic Hamiltonian.
We start with a simpler $su(2)$ non-Hermitian Hamiltonian and construct metrics and isospectral Hermitian counterparts. Then, the analysis become more systematic as we investigate the use of metrics with exponents which are linear in the generators of the referred algebra, first to study linear Hamiltonians and later to consider the inclusion of quadratic terms. The following step is to explore more complicated similarity transformations capable of mapping non-Hermitian into Hermitian systems. In particular, we use exponents of the Dyson operator generated by elements of the enveloping algebra $U(su(2))$.
Finally, the aforementioned results are compared with an explicit diagonalisation of Hamiltonians, thus allowing to make a correspondence between the pseudo-quasi-Hermitian approach and the Swanson method.

\section{Symmetric Non-Hermitian Hamiltonians}

This section is devoted to exploring the symmetries of Hamiltonians expressed in terms of certain operators. If the latter are generators of a particular Lie algebra, we investigate the transformations which leave the associated commutation relations invariant. For convenience, the algebras $su(1,1)$ and $su(2)$ are now studied, with respect to their automorphisms, in a unified way.

\subsection{A general framework for the invariant Hamiltonians}\label{subframe}

In \cite{afkk} an algebraic approach was implemented in order to find metrics for the quadratic generalisation of the so-called Swanson Hamiltonian. There, the method employed is based on the symmetry $su(1,1)$. However, one can easily adapt the same framework to $su(2)$ algebra, whose generators can be realised in such a way that Bose-Einstein condensates can be described \cite{ptbec}.
Despite being very different from a representation theory point of view (the latter is compact, admiting finite dimensional representations, whereas the former noncompact), both algebras have a very similar structure, with three generators, $K_0, K_\pm$ and $L_0, L_\pm$. In this section we will describe them in a unified way as much as possible and for that matter we introduce the $M$ operators. Consider certain operators  $M_0, M_1, M_2$ satisfying the commutation relations
\begin{equation}
 [M_1, M_2]={i} \sigma M_0 , \;\;\;\;\;\;\;\; [M_0, M_1]={i}
 M_2,  \;\;\;\;\;\;\;\; [M_2, M_0]={i}
 M_1.
 \end{equation}
Then, the operators $M_0, M_\pm = M_1 \pm {i} M_2$ commute
according to
\begin{equation}\label{comm}
[M_0, M_\pm]=\pm M_\pm , \;\;\;\;\;\;\;\; [M_+, M_-] = 2 \sigma
M_0 .
 \end{equation}

 It is clear that the choice $\sigma =1$ corresponds to the $su(2)$-Lie algebra, with generators denoted by $M_i \equiv L_i$, whereas $\sigma =-1$ represents $su(1,1)$, whose generators are taken to be $M_i \equiv K_i$.  There exist different automorphisms which leave these algebras, treated on the same footing, invariant \cite{andreasprivate}:
\begin{equation}\label{auto}
\begin{array}{cccccccc}
\tau _1: & & M_{0} & \rightarrow & M_{0},  & M_{\pm} & \rightarrow & M_{\pm} ,\\
\tau _2: & & M_{0} & \rightarrow & -M_{0}, & M_{\pm} & \rightarrow & -M_{\mp}, \\
\tau _3: & & M_{0} & \rightarrow & -M_{0}, & M_{\pm} & \rightarrow & M_{\mp},
\end{array}
\end{equation}

\ni and compositions of them.
In fact, the mapping $H_i \ra -H_i$
and $E_{\pm\alpha} \ra -E_{\mp\alpha}$, of the Cartan subalgebra
generators and step operators respectively, is an automorphism of
any semi-simple Lie algebra due to the invariance of the root
diagram under $\alpha \ra -\alpha$. In this case it is an inner
automorphism because the inversion is an element of the Weyl
group. The first one is clearly just the identity transformation at the level of the $M$ operators.
However, it may still correspond to nontrivial symmetries, when expressed in terms of physical quantities in a particular representation of the algebra, as will be discussed below.

The invariance of the algebras depicted above can be interpreted physically, in terms of coordinates and momenta, if we use representations in terms of the bosonic operators
\begin{equation}\label{eqn_a_p_x}
 a_j=\frac{\omega_j \,  x_j+{i} p_j}{\sqrt{2\omega_j }} \;\;\;\;\; \textrm{and} \;\;\;\;\;
 a_j^{\dagger}=\frac{\omega_j \,  x_j-{i} p_j}{\sqrt{2\omega _j}},\quad {\rm with }\quad \omega_j\in\mathds{R}.\label{eqn_apx}
\end{equation}
In cases where there is only one kind of boson, the index $j$ may be dropped. Here the bosonic creation and annihilation operators $a^\dag, a$ can be defined as usual in terms of coordinate and momentum coordinates $x, p$ via
(\ref{eqn_apx}). These operators, then, provide one with a representation of
$su(1,1)$ if one takes
\begin{equation}\label{twoboson}
K_{0}=\frac{1}{2}\left(a^{\dag }a+\frac{1}{2}\right),\qquad
K_{+}=\frac{1}{2}a^{\dag }a^{\dag },\qquad
K_{-}=\frac{1}{2}aa .
\end{equation}
In this representation we may realize the automorphisms (\ref{auto}) as
\begin{center}
\begin{equation}\label{auto1}
\begin{array}{cccccccccc}
\tau _1  & : &  a & \rightarrow & -a, & a^{\dagger} & \rightarrow & -a^{\dagger}, & \Longleftrightarrow & \mathcal{PT},  \\
\tau _2  & : & a & \rightarrow & {i} a^{\dagger}, & a^{\dagger} & \rightarrow & {i} a, & \Longleftrightarrow & C, \\
\tau _3 & : & a & \rightarrow & -a^{\dagger}, & a^{\dagger} & \rightarrow & a, & \Longleftrightarrow & \tau_{xp} .
\end{array}
\end{equation}
\end{center}

Thus, $\tau_1$ is the usual $\mathcal{PT}$ transformation, an anti-linear operation reversing time and space coordinates: $x  \rightarrow  -x,  \; p  \rightarrow  p,  \; {i}  \rightarrow - {i}$. Note that although this transformation can be achieved by the identity operator in (\ref{auto}), when writing the $M$ operators in terms of bosonic operators, a nontrivial transformation associated to $\tau_1$ may be constructed with the use of $a, a^\dag$. The automorphism $\tau_1$ can be taken to be just the identity also in (\ref{eqn_apx}) but the possibility of using the equally valid $\mathcal{PT}$ operation is more interesting.

A different transformation is introduced by $\tau_2$, which changes the sign of the energy: $\omega  \rightarrow  -\omega$. Having in mind the Dirac sea structure, one can interpret this operation as charge conjugation relating particles to antiparticles. One should not mistake this $C$ with the one used in \cite{sense}, and denoted $\mathcal{C}$, for a definition of the scalar product in the Hilbert space. The \cpt operation can then be seen as composition of the first and second symmetries: $C\mathcal{PT} \equiv \tau_1 \tau_2 : K_0 \rightarrow -K_0, K_\pm \rightarrow -K_\mp$, a combination of charge conjugation with parity and time reversal transformations.
The last transformation $\tau_3$, on the other hand, is a symmetry in the phase space, intertwining coordinate and momentum:  $x  \rightarrow  \frac{{i}}{\omega} p, \;  p \rightarrow {i} \omega x$.

One may also represent the same algebra by means of to two distinguishable bosons
\begin{equation}\label{novaesrep}
K_0 = \frac{1}{2} \( a_1^\dag a_1 + a_2^\dag a_2 + 1\) ,
\;\;\;\;\;\; K_+ = a_1^\dag a_2^\dag , \;\;\;\;\;\; K_- = a_1 a_2.
\end{equation}
The automorphism can then be described through
\begin{equation}\label{auto2}
\begin{array}{cccccccccccccc}
\tau _1  & : & a_1  & \rightarrow & -a_1, & a_1^{\dagger} & \rightarrow & -a_1^{\dagger}, \\
 & & a_2 & \rightarrow & -a_2, & a_2^{\dagger} & \rightarrow & -a_2^{\dagger} ,\\
\tau _2  & : & a_1 & \rightarrow & a_2^{\dagger}, & a_1^{\dagger } & \rightarrow & a_2, \\
 & & a_2 & \rightarrow & -a_1^{\dagger}, & a_2^{\dagger } & \rightarrow & -a_1, \\
\tau _3 & : & a_1 & \rightarrow & a_2^{\dagger}, & a_1^{\dagger} & \rightarrow & -a_2, \\
 & & a_2 & \rightarrow & a_1^{\dagger}, & a_2^{\dagger } & \rightarrow & -a_1 .
\end{array}
\end{equation}

Again $\tau_1$  is the usual antilinear $\mathcal{PT}$-symmetry transformation, reversing coordinate and momenta of the same kind of bosons:
${i} \rightarrow -{i}, x_i \rightarrow - x_i, p_i \rightarrow p_i$. In the coordinates and momenta representation, $\tau_2$ can be achieved by considering
$\omega_{1}=\omega _{2}=\omega $ together with ${i} \rightarrow -{i},  x_1 \rightarrow x_2, p_1 \rightarrow p_2 , x_2 \rightarrow-x_1, p_2 \rightarrow -p_1$. Also, the third automorphism, $\tau_3$, relates position and momentum; $x_i \rightarrow \pm \frac{{i}}{\omega_i}p_i$.

Another similarity between the algebras $su(1,1)$ and $su(2)$ is that the former can also be realised with the Schwinger Dyson transformation \cite{Schwinger}
\begin{equation}
L_{0}=\frac{1}{2}(a_{1}^{\dag }a_{1}-a_{2}^{\dag }a_{2}),\qquad
L_{+}=a_{1}^{\dag }a_{2},\qquad L_{-}=a_{2}^{\dag }a_{1},
\end{equation}
that maps the $su(2)$ algebra to a two-dimensional Heisenberg-Weyl algebra with generators $ a_{1,2}^\dagger$ and $ a_{1,2}$. The most prominent example of this realisation are ensembles of two-state boson systems, where $ a_j^\dagger$ creates a particle in state $j$, and $ a_j$ annihilates one of them. Alternative realisations are given by two-dimensional single particle or one-dimensional two-particle quantum systems whose position and momentum operators $ x_j$ and $ p_j$ are defined according to (\ref{eqn_a_p_x}).

In this representation the automorphisms satisfying (\ref{auto}) behave as
\begin{equation}\label{auto3}
\begin{array}{lllllllll}
\tau _1& : & a_1 & \rightarrow & -a_1, & a_1^{\dagger} & \rightarrow & -a_1^{\dagger}, \\
& & a_2 & \rightarrow & -a_2, & a_2^{\dagger} & \rightarrow & -a_2^{\dagger},  \\
\tau _2 & : & a_1 &\rightarrow & - {i} a_2, & a_1^{\dagger} & \rightarrow & {i} a_2^\dagger, \\
& & a_2 & \rightarrow & {i} a_1, & a_2^{\dagger} & \rightarrow & -{i} a_1^{\dagger}, \\
\tau _3 & : & a_1 & \rightarrow & a_2, & a_1^{\dagger} & \rightarrow & a_2^{\dagger}, \\
& & a_2  & \rightarrow & a_1, & a_2^{\dagger} & \rightarrow & a_1^{\dagger}.
\end{array}
\end{equation}

In the ensemble of two-state systems realisation, the first symmetry is basically the identity, whereas the second and the third automorphisms correspond to interchanging the two states and can thus naturally be associated with parity. Associating the Heisenberg-Weyl algebra with position and momentum operators via (\ref{eqn_a_p_x}) however, the automorphisms can be interpreted quite differently.

Now, $\tau_1$ still corresponds to a \pt-symmetry (${i} \rightarrow -{i}, x_i \rightarrow - x_i, p_i \rightarrow p_i$) while the second automorphism $\tau_2$
rotates the phase space by $\frac{\pi}{2}$ in complex plane, $w_1 \leftrightarrow w_2, x_1 \leftrightarrow \pm {i} x_2, p_1 \leftrightarrow \pm {i} p_2 $.
Finally, $\tau_3$ just intertwines the particle's labels: $w_1 \leftrightarrow w_2, x_1 \leftrightarrow x_2, x_1 \leftrightarrow x_2$.
This last one corresponds to the symmetry imposed in the system described in \cite{ptbec}. This realisation might be interpreted as involving bosons
of different species transmuting into one another, bosons of the same kind but in different positions swapping their coordinates, or to describe a two-dimensional problem according to (\ref{eqn_a_p_x}).

Although the algebras $su(2)$ and $su(1,1)$ are not equivalent (since the former corresponds to a compact group and allows for a finite dimensional representation while the latter arises from a noncompact group and has only infinite dimensional representations),  it is interesting to note that by applying $L_0 \rightarrow -L_0 , \;\; L_\pm \rightarrow \mp L_\mp$, which of course is not an automorphism, one can construct combinations which satisfy the commutation relation of the $K$s, i.e.  (\ref{comm}) with $\sigma =-1$.

The framework described in this section is general and can be applied to other families of algebras in order to construct different anti-linear symmetries of the Hamiltonian. These symmetries, as we know, have an important role in selecting systems which potentially have real spectra and that consequently are interesting from a physical point of view. Moreover, it contains the whole class of Hamiltonians studied in this manuscript as a subcase and other families can be constructed similarly.

\subsection{Non-Hermitian su(2) Hamiltonians with an anti-linear symmetry}

Here we are interested in generic non-Hermitian Hamiltonians expressed in terms of $su(2)$ generators, that is, Hamiltonians which are composed of angular momentum operators $ L_x \equiv  L_1,\,  L_y\equiv  L_2,\,  L_z\equiv  L_0$, that fulfill the relations $[L_x, L_y]={i} \epsilon_{x,y,z} L_z $ with $ \epsilon_{x,y,z}$ the Levi-Civita antisymmetric symbol. Alternatively one can use $ L_0= L_z,\,  L_\pm =  L_x \pm {i}  L_y$,  which commute as
\begin{equation}\label{comm}
[ L_0,  L_\pm]=\pm  L_\pm , \;\;\;\;\;\;\;\; [ L_+,  L_-] = 2  L_0.
\end{equation}
The physical importance of $su(2)$ Hamiltonians goes far beyond the description of actual angular momentum systems such as, e.g., nuclear spins \cite{Lipkin}. The basis for many other important physical realisations of quantum $su(2)$ systems is in terms of the bosonic operators (\ref{eqn_a_p_x}). In Hermitian quantum mechanics quadratic $su(2)$ Hamiltonians are of importance in fields such as nuclear physics and Bose-Einstein condensates \cite{ptbec}. These models are also very interesting due to the possibility of using semi-classical analysis in order to construct the scalar product of the non-Hermitian quantum system by relating the quantum Dyson operator with the Kahler metric for the corresponding classical Hamiltonian systems \cite{ptkahler}. If possible, this could bring in an alternative approach to constructing metrics. Here we are interested in non-Hermitian generalisations of $su(2)$ Hamiltonian systems.

The most general complex Hamiltonian quadratic in the $su(2)$ operators belongs to an $18$ parameter family
\be\label{eqn_gen_H_quad}
 H = \sum_{j=0}^{2} (\alpha_j+{i} \beta_j) L_j+\sum_{0 \leq j < k \leq 2}(\alpha_{jk}+{i}\beta_{jk}) L_j L_k.
\ee
Any analytical function of $ L_j$ commutes with $ L^2= L_x^2+ L_y^2+ L_z^2 = L_0^2 +\frac{1}{2}(L_+L_- + L_-L_+)$. Thus the representation of such a function in the standard angular momentum basis has  a block diagonal structure, where the $n$-th block is of size $n\times n$. For the Hamiltonian (\ref{eqn_gen_H_quad}), in particular, each block has a band structure.
These Hamiltonians, however, are in general not pseudo-quasi-Hermitian. In what follows we will impose an additional anti-linear symmetry, which decreases the number of free parameters and allows for possible pseudo-quasi-Hermiticity.

Below we comment on possible physical realisations and interpretations of the three automorphisms of the $su(2)$ algebra.
For the moment we shall investigate Hamiltonians fulfilling one particular symmetry as an example, $\tau_3$ (\ref{auto}). Other cases can be treated analogously. Let us then consider Hamiltonians invariant under the third automorphism $\tau_3$ combined with complex conjugation, that is
\be\label{eqn_symm}
 L_0\to - L_0,\quad  L_{\pm}\to  L_{\mp},\quad {i} \to-{i},
\ee
or expressed in the angular momentum components
\be
 L_x\to  L_x,\quad  L_y\to  L_y,\quad  L_z\to - L_z,\quad {i} \to-{i}.
\ee
Requiring the quadratic $su(2)$ Hamiltonian (\ref{eqn_gen_H_quad}) to be invariant under this transformation reduces the number of free parameters from $18$ to $9$. Explicitly, the Hamiltonians under investiagtion in the present paper are of the form
\begin{eqnarray}\label{llgeneral}
\nonumber
\fl
\qquad H &=& {i} \beta_0 L_0 + (\alpha_+ +
{i} \beta_+)L_+ + (\alpha_+ - {i} \beta_+)L_- + \\
\fl &+& \alpha_{00} L_0^2 + (\alpha_{+0} + {i} \beta_{+0})L_+ L_0 -
(\alpha_{+0} - {i} \beta_{+0})L_- L_0 +\\
\nonumber \fl &+&  (\alpha_{++} + {i} \beta_{++})L_+^2 + (\alpha_{++} - {i} \beta_{++})L_-^2 + \alpha_{+-}(L_+ L_- + L_- L_+),
\end{eqnarray}

\ni being also expressible in terms of $L_z=L_0, \; L_x = \frac{1}{2}(L_+ + L_-), \; L_y = \frac{1}{2 {i}}(L_+ - L_-)$:
\begin{eqnarray}
\nonumber H & = &  {i} \varsigma_z L_z + \varsigma_x L_x + \varsigma_y L_y + \varsigma_{zz} L_z^2 +\varsigma_{xx} L_x^2 + \varsigma_{yy} L_y^2 + \\
&+& \varsigma_{xy} (L_x L_y + L_y L_x) + {i} \varsigma_{xz} L_x L_z + {i} \varsigma_{yz} L_y L_z .
\end{eqnarray}
The exact matching between the two equivalent systems above is simply achieved with
\begin{eqnarray}\label{equivaa}
\nonumber \varsigma_x=2\alpha_+, & \varsigma_y=-2\beta_+, & \varsigma_z=\beta_0, \\
\varsigma_{xx}=2(\alpha_{+-}+\alpha_{++}), \;\;\; & \varsigma_{yy}=2(\alpha_{+-}-\alpha_{++}), \;\;\; & \varsigma_{zz}=\alpha_{00}, \\
\nonumber \varsigma_{xy}=-2\beta_{++}, & \varsigma_{xz}=2\beta_{+0}, & \varsigma_{yz}=2\alpha_{+0}.
\end{eqnarray}

In the linear part of the Hamiltonian (\ref{llgeneral}), the non-Hermiticity  comes from the $L_0$ term, differently from the $su(1,1)$ analogue. In that situation, the $K_0$ term would not break Hermiticity but the corresponding `ladder' operators $K_\pm$ would, unlike here. Also, the fact that the Hermitian conjugate of $L_\pm L_0$ is not $L_\mp L_0$, but rather $L_0L_\mp$, contributes for the breaking of Hermiticity in the quadratic part of (\ref{llgeneral}). Clearly we can change the orders of these operators and gain extra terms in the direction of $L_\pm$.

Despite the fact of having $H \neq H^\dag$, we might still avoid complex eigenvalues.
If the anti-linear symmetry of a Hamiltonian of this class is unbroken, it leads to a purely real spectrum and the existence of a Hermitian counterpart. For the construction of a subclass of non-Hermitian systems, the requirement of unbroken symmetry, however, is of little practical use, as it basically requires the explicit knowledge of the eigenfunctions. In what follows, we shall use a different approach to construct a subclass of non-Hermitian Hamiltonians, using Lie algebraic properties and an explicit ansatz for the $\eta$ operator. This will allow us to construct not the whole, but a large, class of non-trivial pseudo-quasi-Hermitian Hamiltonians of the type (\ref{llgeneral}).

\section{Metrics and Hermitian counterparts}\label{secmet}

Before we proceed to the general quadratic case, let us first illustrate the basic ideas for the analytically solvable case. In the study of an $N$ particle two-mode Hubbard system, a natural Hamiltonian used to describe a Bose-Einstein condensate in a double-well potential emerges, which was made non-Hermitian in \cite{ptbec} by the introduction of an imaginary $L_z$ term. It is instructive to represent the Hamiltonian in the components of the angular momentum operators as
\begin{equation}\label{hameva}
H= {i} \varsigma_z L_z +\varsigma_x L_x + \varsigma_{zz} L_z^2.
\end{equation}

It was observed that in the vanishing interaction limit, with $a_{zz} \rightarrow 0$, the Hamiltonian can be expressed as
\begin{equation}\label{HtoL0}
H = -\sqrt{\varsigma_x^2-\varsigma_z^2} e^{-\textrm{arctanh}\left( \frac{\varsigma_x}{\varsigma_z} \right)L_y }L_z  e^{\textrm{arctanh}\left( \frac{\varsigma_x}{\varsigma_z} \right)L_y },
\end{equation}
indicating that it is in the same similarity class as the Hermitian Hamiltonian $h = -\sqrt{\varsigma_x^2-\varsigma_z^2} L_z$, with the operator $\eta = e^{-\textrm{arctanh}\left( \frac{\varsigma_x}{\varsigma_z} \right)L_y} $, thus, acting as a Dyson map.  The $ \eta$ operator diverges at $| \varsigma_z |=| \varsigma_x |$. At this so-called exceptional point \cite{EP1, EP2}, the Hamiltonian is not diagonalisable, but has a Jordan block structure, where all eigenvalues and their corresponding eigenvectors degenerate. That is, at this point there is no equivalent Hermitian operator. For values of $|\varsigma_z|$ above this critical value, $ \eta$ diagonalises the Hamiltonian again. For the interacting Hamiltonian, with a quadratic part, a metric operator of this simple form could not be found, though.


While only some special systems can be solved via a direct diagonalisation, an $\eta$ operator of a simple exponential form that maps the Hamiltonian to a Hermitian, but not diagonal, counterpart can be found in certain circumstances. For this purpose, an explicit ansatz for $ \eta$ is used to act adjointly on the Hamiltonian and the result is demanded to be Hermitian. This then yields constraints for the parameters of the Hamiltonian and of the $ \eta$ operator. Here we will investigate $ \eta$ operators whose exponent is linear in the $su(2)$ operators and respect the anti-linear symmetry of the Hamiltonian.


We start by analysing the linear Swanson-like model and then the more general families of quadratic Hamiltonians in terms of a metric with linear exponent with respect to the generators of the algebra. Later we introduce a metric with purely quadratic terms and finally we compare the similarity transformation underlying the pseudo-quasi-Hermitian approach with the Swanson method of diagonalization \cite{swanson} and establish a connection between them.

\subsection{Conditions for a linear exponent metric}\label{sec_met}

To find the whole class of quadratic Hamiltonians that can be mapped to a Hermitian counterpart by a simple $\eta$ with an exponent linear in $L_j$, an explicit ansatz for the Dyson operator is applied to the Hamiltonian and the result is demanded to be Hermitian. This then yields constraints for the parameters of the Hamiltonian and of the $\eta$  operator. In what follows we will investigate Dyson operators whose exponent is linear in the $su(2)$ generators, that is, we make the ansatz
\begin{equation}\label{metansatz}
\eta=\exp{\left[2\left(\lambda_{0}L_{0}+\lambda_{+}L_{+}+\lambda_{-}L_{-}\right)\right]},
\end{equation}
\ni in terms the three parameters $\lambda_{0,\pm}$.
If we want the above transformation $\eta$ not to break the eventual anti-linear symmetry of a Hermitian counterpart wave-function $| \phi \rangle$ for the non-Hermitian eigenstate $| \Phi \rangle = \eta^{-1}| \phi \rangle$, we should impose $\eta$ to be symmetric as well, implying
$\lambda_0 = {i} \Gamma_0, ~\lambda_+ =  \lambda + {i} \Gamma, ~ \lambda_- = \lambda - {i} \Gamma$:
\begin{eqnarray}\label{bigeta}
 \eta &=& \exp{\left[2\({i} \Gamma_0  L_0 +
(\lambda + {i}\Gamma) L_+ + (\lambda - {i} \Gamma) L_- \) \right]}\nonumber\\
&=&\exp{\(\lambda_x  L_x + \lambda_y  L_y + {i}\lambda_z  L_z\)}\label{eqn_eta_ansatz1},
\end{eqnarray}
with $\lambda_x = 4\lambda,~ \lambda_y =-4 \Gamma, ~\lambda_z = 2\Gamma_0 \in\mathds{R}$.
The operator can be made Hermitian by choosing $\Gamma_0=0$ and its positivity is guaranteed by the condition $\theta^2 =  4(\lambda^2 + \Gamma^2) -\Gamma_0^2 \equiv \frac{1}{4}\phi^2 = \frac{1}{4}\left( \lambda_x^2 + \lambda_y^2 -\lambda_z^2 \right)\geq 0$. Due to the linearity of the exponent with respect to the $su(2)$ generatores, it conserves the powers of the $su(2)$ operators appearing in $ H$.

The adjoint action of the $\eta$ operator (\ref{metansatz}) on the $L$ generators can be characterised by the matrix in the following equation:
\begin{equation}\label{simmatrix}
\left(
\begin{array}{c}
\tilde{L}_{0} \\
\tilde{L}_{+} \\
\tilde{L}_{-}
\end{array}
\right) = \left(
\begin{array}{ccc}
b_{00} & b_{0+} & b_{0-} \\
b_{+0} & b_{++} & b_{+-} \\
b_{-0} & b_{-+} & b_{--}
\end{array}
\right) \left(
\begin{array}{c}
L_{0} \\
L_{+} \\
L_{-}
\end{array}
\right),
\end{equation}
\ni whose determinant is simply unity and elements are given by

\begin{equation}
\begin{array}{ll}
\fl
b_{00}=1+8\lambda_+ \lambda_- \( \frac{\sinh{\theta}}{\theta} \)^2, ~
b_{\pm \mp}=-\( 2 \lambda_\mp \frac{\sinh{\theta}}{\theta} \)^2, ~
b_{\pm \pm} = \( \cosh{\theta} \pm \lambda_0
\frac{\sinh{\theta}}{\theta} \)^2, \\
\ \\
\fl
b_{0 \pm} = \mp 2 \lambda_\pm \frac{\sinh{\theta}}{\theta}\(
\cosh{\theta} \pm \lambda_0
\frac{\sinh{\theta}}{\theta} \), ~~~
b_{\pm 0} = \mp 4 \lambda_\mp \frac{\sinh{\theta}}{\theta}\(
\cosh{\theta} \pm \lambda_0
\frac{\sinh{\theta}}{\theta} \) .
\end{array}
\end{equation}

In some circumstances it is more convenient to use the metric ordered according to a Gauss decomposition
\footnote{This form is particularly interesting when acting on a highest weight state $| \mu \rangle$ such that  $L_+| \mu \rangle =0$.\\}, such as
\begin{eqnarray}\label{etagauss}
\eta = e^{2\kappa_- L_-}e^{2\kappa_0 L_0}e^{2\kappa_+ L_+}.
\end{eqnarray}
In this case the parameters relate easily to the ones in  (\ref{metansatz})  as
\footnote{Using the results in \cite{afkk},  the expressions above can be more easily calculated by noting
that the they can be rewritten using a minor adjustment to reproduce now the algebra $su(2)$ from $su(1,1)$ generators
according to $K_0 \rightarrow -L_0, K_+ \rightarrow L_-, K_- \rightarrow - L_+ $. The comparison with the relations obtained previously for
$su(1,1)$  can be used to obtain the expressions above according to $\lambda_0 \ra -\lambda_0, \lambda_+ \ra \lambda_-, \lambda_- \ra -\lambda_+$ and $\kappa_0 \ra -\kappa_0, \kappa_+ \ra \kappa_-, \kappa_- \ra -\kappa_+$.}
\begin{equation}\label{gaussdec}
\fl
\kappa_0 = \log{\( \cosh{\theta} + \lambda_0 \frac{\sinh{\theta}}{\theta}\)},
\kappa_+ =\frac{2\lambda_+\frac{\sinh{\theta}}{\theta}}{\cosh{\theta}+\lambda_0 \frac{\sinh{\theta}}{\theta}},
\kappa_- =\frac{2\lambda_-\frac{\sinh{\theta}}{\theta}}{\cosh{\theta}+\lambda_0 \frac{\sinh{\theta}}{\theta}}.
\end{equation}

We have already discussed that the non-Hermiticity of the Hamiltonian (\ref{llgeneral}) results from the $L_0$ term and from $L_+L_0$ and $L_0L_-$.
However, when we impose $h = h^\dag$ in (\ref{dysonherm}), we notice that that the adjoints of latter two terms can be recast in terms of the original operators after a simple commutation, introducing terms proportional to $L_+$ and $L_-$. Therefore, all terms are Hermitian except for the linear components and there are three equations originated from these terms which need to be solved.
After manipulation of the resulting constraints and the use of standard trigonometric identities, one can notice that they can be put in the form of fourth degree polynomial equations. For practical purposes, it is convenient to introduce the quantity $Y \equiv \frac{\tanh{\(\theta\)}}{\theta} $, which is just an abbreviation for a natural combination appearing in the constraints.  Because $Y$ depends on $\theta$ in a transcendental way, solving for $\theta$ explicitly remains a challenge. It is important to note that the quantity $Y$ has no deeper physical meaning, with $\theta$ (or ultimately, $\lambda, \Gamma, \Gamma_0$) being the important variable (or combination of variables).
These constraints are found to be of the form

\begin{eqnarray}\label{constraints}
\xi_0^{(i)} + \xi_1^{(i)} Y + \xi_2^{(i)} Y^2 + \xi_3^{(i)} Y^3 + \xi_4^{(i)} Y^4 =0
\end{eqnarray}
with $i=1,2,3$ and coefficients given by the expressions bellow:

\begin{eqnarray}\label{setconst}
\fl
\nonumber
\xi_0^{(1)} = \beta_0, \qquad \qquad
\xi_1^{(1)} = 4 \Gamma  \left(2 \alpha _+-\alpha _{+0}\right)-4 \lambda  \left(2 \beta _+-\beta _{+0}\right), \\
\fl
\nonumber
\xi_2^{(1)} = 2 \Gamma _0 \left(2 \lambda  \alpha _{+0}+2 \Gamma  \beta _{+0}-4 \alpha _+ \lambda -4 \beta _+ \Gamma +\beta _0 \Gamma _0\right), \\
\fl
\nonumber
\xi_3^{(1)} = 4 \left(4 \left(\Gamma ^2+\lambda ^2\right)-\Gamma _0^2\right) \left(\Gamma  \left(\alpha _{+0}-2 \alpha _+\right)-\lambda  \beta _{+0}+2 \beta _+ \lambda
   \right), \\
\fl
\nonumber
\xi_3^{(1)} =\left(\Gamma _0^2-4 \left(\Gamma ^2+\lambda ^2\right)\right) \left(4 \Gamma _0 \left(\lambda  \alpha _{p,0}+\Gamma  \beta _{p,0}-2 \alpha _+ \lambda -2 \beta _+
   \Gamma \right)+\beta _0 \left(4 \left(\Gamma ^2+\lambda ^2\right)+\Gamma _0^2\right)\right),  \\
\fl
\nonumber
\\
\fl
\nonumber
\xi_0^{(2)} = \alpha_{+0}, \qquad \qquad
\xi_1^{(2)} = -2 \left(\Gamma _0 \beta _{+0}+2 \lambda  \left(\alpha _{00}-2 \alpha _{+-}+2 \alpha _{++}\right)+4 \Gamma  \beta _{++}\right), \\
\fl
\nonumber
\xi_2^{(2)} = 4 \left(6 \lambda  \left(\lambda  \alpha _{+0}+\Gamma  \beta _{+0}\right)+\Gamma _0 \left(\Gamma  \left(\alpha _{00}-2 \left(\alpha _{+-}+3 \alpha
   _{++}\right)\right)+6 \lambda  \beta _{++}\right)\right), \\
\fl
\nonumber
\xi_3^{(2)} = -4 \lambda  \alpha _{00} \left(4 \left(\Gamma ^2+\lambda ^2\right)+\Gamma _0^2\right)-24 \Gamma _0 \left(\Gamma ^2+\lambda ^2\right) \beta _{+0}-2 \Gamma _0^3
   \beta _{+0} + \\
   \fl 
   \nonumber 
   \qquad + 32 \left(\Gamma ^2 \lambda  \left(\alpha _{+-}+3 \alpha _{++}\right)+\lambda ^3 \left(\alpha _{+-}-\alpha
   _{++}\right)+\Gamma ^3 \beta _{++}-3 \Gamma  \lambda ^2 \beta _{++}\right) \\
   \fl
   \nonumber
   \qquad 
   +8 \Gamma _0^2 \left(\lambda  \alpha _{+-}+3 \lambda 
   \alpha _{++}+3 \Gamma  \beta _{++}\right), \\
\fl
\nonumber
\xi_4^{(2)} = 16 \left(\Gamma ^2+\lambda ^2\right) \left((\lambda -\Gamma ) (\Gamma +\lambda ) \alpha _{+0}+2 \Gamma  \lambda  \beta _{+0}\right)+24 \Gamma  \Gamma _0^2
   \left(\lambda  \beta _{+0}-\Gamma  \alpha _{+0}\right)+\Gamma _0^4 \left(-\alpha _{+0}\right) \\
   \fl
   \nonumber
   \qquad 
   +16 \Gamma _0 \left(\Gamma ^3 \left(\alpha _{00}-2 \alpha
   _{+-}+2 \alpha _{++}\right)+\Gamma  \lambda ^2 \left(\alpha _{00}-2 \left(\alpha _{+-}+3 \alpha _{++}\right)\right)+
   \right. \\
   \fl
   \nonumber
   \qquad 
   \left.
   +2\left(\lambda ^3-3 \Gamma ^2 \lambda \right) \beta _{++}\right)
   +4 \Gamma _0^3 \left(\Gamma  \left(\alpha _{00}-2 \alpha _{+-}+2 \alpha
   _{++}\right)-2 \lambda  \beta _{++}\right), \\
\fl
\nonumber
 \\
\fl
\nonumber
\xi_0^{(3)} =\beta_{+0}, \qquad \qquad
\xi_1^{(3)} = -4 \Gamma  \alpha _{00}+2 \Gamma _0 \alpha _{+0}+8 \Gamma  \left(\alpha _{+-}+\alpha _{++}\right)-8 \lambda  \beta _{++}, \\
\fl
\nonumber
\xi_2^{(3)} = 24 \Gamma  \left(\lambda  \alpha _{+0}+\Gamma  \beta _{+0}\right)-4 \Gamma _0 \left(\lambda  \left(\alpha _{00}-2 \alpha _{+-}+6 \alpha
   _{++}\right)+6 \Gamma  \beta _{++}\right), \\
\fl
\nonumber
\xi_3^{(3)} = 2 \left(12 \Gamma _0 \left(\Gamma ^2+\lambda ^2\right) \alpha _{+0}+\Gamma _0^3 \alpha _{+0}-8 \Gamma ^3 \left(\alpha _{00}-2 \left(\alpha
   _{+-}+\alpha _{++}\right)\right) \right. \\
   \fl
   \nonumber
   \qquad 
   \left.
   -8 \Gamma  \lambda ^2 \left(\alpha _{00}-2 \alpha _{+-}+6 \alpha _{++}\right)-2 \Gamma _0^2
   \Gamma  \left(\alpha _{00}-2 \alpha _{+-}+6 \alpha _{++}\right) \right.\\
   \fl
   \nonumber
   \qquad
   \left.
   +4 \lambda  \left(3 \left(\Gamma _0^2-4 \Gamma ^2\right)+4 \lambda ^2\right)
   \beta _{++}\right), \\
\fl
\nonumber
\xi_4^{(3)} = 4 \left(8 \Gamma  \lambda  \left(\Gamma ^2+\lambda ^2\right) \alpha _{+0}+6 \Gamma  \Gamma _0^2 \lambda  \alpha _{+0}-4 \Gamma _0 \left(\Gamma ^2 \lambda 
   \left(\alpha _{00}-2 \alpha _{+-}+6 \alpha _{++}\right) \right. \right. \\
   \fl
   \nonumber
   \qquad
   \left. \left.
   +\lambda ^3 \left(\alpha _{00}-2 \left(\alpha _{+-}+\alpha
   _{++}\right)\right)+2 \Gamma ^3 \beta _{++}-6 \Gamma  \lambda ^2 \beta _{++}\right) \right. \\
   \fl
   \nonumber
   \qquad 
   \left.
   +\Gamma _0^3 \left(-\lambda  \alpha _{00}+2 \lambda 
   \left(\alpha _{+-}+\alpha _{++}\right)+2 \Gamma  \beta _{++}\right)\right)-\left(24 \Gamma _0^2 \lambda ^2+\Gamma _0^4+16
   \lambda ^4-16 \Gamma ^4 \right) \beta _{+0}.
\end{eqnarray}
\normalsize

\vspace{0.25cm}

Therefore, our strategy transformed the problem of determining the Dyson operator, the metric and the Hermitian counterpart associated to equation (\ref{llgeneral}), ultimately, into that of solving the above set of coupled transcendental equations for the parameters $\alpha$ and $\beta$ in the Hamiltonian, as well as $\lambda, \Gamma, \Gamma_0$ introduced by the similarity transformation. Despite having completely different starting points and little in common, this technique can be seen as having the same spirit as the Bethe ansatz approach for integrable systems, for which the complete diagonalization of a problem is reduced to a set of coupled equations for the spectral parameters \cite{bethe}.

Once the Dyson operator is obtained the metric $\rho = \eta^\dag \eta$ can then be constructed as well as the Hermitian isospectral counterpart  according to (\ref{dysonherm}), providing us with

\begin{eqnarray}\label{transllgeneral}
\fl
\qquad h &=&  -2 A_{+-} L_0 + A_{00} L_0^2 + 2A_{+-} L_+L_-  + \\
\fl \nonumber
&+& \left((A_+ + A_{0+} + i(B_+ + B_{0+}) \right) L_+  + \left((A_+ + A_{0+} - i(B_+ + B_{0+}) \right) L_- +
\\
\fl \nonumber
&+& (A_{++} + i B_{++}) L_+^2 + (A_{++} - i B_{++}) L_-^2 + \\
\fl \nonumber
&+& \left( A_{0+}+A_{+0} + i(B_{0+}+B_{+0}) \right)L_+ L_0 +
\left( A_{0+}+A_{+0} - i(B_{0+}+B_{+0}) \right)L_0 L_-,
\end{eqnarray}

\vspace{0.2cm}

\ni where each line in the right hand side of the equation above is Hermitian, given the coefficients
$A_i=\cosh^2\theta \; \bar{A}_i,
~ A_{ij}=\cosh^4\theta \; \bar{A}_{ij},
~ B_j=\cosh^2\theta \; \bar{B}_j,
~ B_{ij}=\cosh^4\theta \; \bar{B}_{ij}$ are real. In fact they can be expressed as

\begin{eqnarray}
\fl \nonumber
\bar{A}_+ &=& \alpha_+ \left(1-Y^2 \left(-4 \Gamma ^2+\Gamma _0^2+4 \lambda ^2\right)\right)+2 Y \left(\beta_0 \left(\Gamma +\Gamma_0 \lambda  Y\right)-\beta _+
   \left(\Gamma_0+4 \Gamma  \lambda  Y\right)\right), \\
\fl \nonumber
\bar{A}_{00} &=& \alpha _{00} \left(Y^2 \left(4 \left(\Gamma ^2+\lambda ^2\right)+\Gamma _0^2\right)+1\right)^2
-8 Y \left(\Gamma  \beta _{+0}+\Gamma _0 Y^3 \left(4 \Gamma _0 \left(\Gamma ^2 \left(\alpha _{+-}-\alpha _{++}\right) \right.\right. \right. \\
\fl
\nonumber
&+&
\left. \left. \left.
\lambda ^2 \left(\alpha _{+-}+\alpha _{++}\right)+2 \Gamma  \lambda  \beta
   _{++}\right)-\lambda  \left(4 \left(\Gamma ^2+\lambda ^2\right)+\Gamma _0^2\right) \beta _{+0}\right) \right. \\
   \fl
   \nonumber
   &+&
   \left. 
   Y \left(-\Gamma _0 \lambda  \beta _{+0}+4
   \left(\Gamma ^2+\lambda ^2\right) \alpha _{+-}+4 \Gamma ^2 \alpha _{++}-4 \lambda ^2 \alpha _{++}-8 \Gamma  \lambda  \beta
   _{++}\right) \right. \\
   \fl
   \nonumber
   &+&
   \left.
   Y^2 \left(\Gamma  \left(4 \left(\Gamma ^2+\lambda ^2\right)+\Gamma _0^2\right) \beta _{+0}+8 \Gamma _0 \left((\lambda -\Gamma ) (\Gamma
   +\lambda ) \beta _{++}-2 \Gamma  \lambda  \alpha _{++}\right)\right) \right. \\
   \fl
   \nonumber
   &+&
   \left.
  \alpha _{+0} \left(\lambda +\Gamma  \Gamma _0 Y\right) \left(Y^2 \left(4
   \left(\Gamma ^2+\lambda ^2\right)+\Gamma _0^2\right)+1\right)\right), \\
\fl \nonumber
\bar{A}_{+-} &=& 2 Y \left(Y \left(\Gamma  \Gamma _0 \left(\alpha _{+0} \left(4 Y^2 \left(\Gamma ^2+\lambda ^2\right)+1\right)-16 \lambda  Y \alpha _{++}\right) \right. \right. \\
\fl
\nonumber
&+&
\left. \left.
\Gamma_0^2 Y \left(\lambda  \left(\alpha _{+0}-2 \lambda  Y \alpha _{00}+4 \lambda  Y \alpha _{++}\right)-2 \Gamma ^2 Y \left(\alpha _{00}+2 \alpha _{++}\right)\right) \right.\right.\\
\fl
\nonumber
&-&
\left.\left. 2 \Gamma ^2 \left(\alpha _{00}-2 \left(\lambda  Y \alpha _{+0}+\alpha _{++}\right)\right)+\Gamma  \Gamma _0^3 Y^2 \alpha_{+0}-8 \beta _{++} \left(\lambda +\Gamma  \Gamma _0 Y\right) \left(\Gamma -\Gamma _0 \lambda  Y\right)\right) \right.  \\
\fl
\nonumber
&+&
\left.
\beta _{+0} \left(\Gamma -\Gamma _0
   \lambda  Y\right) \left(Y^2 \left(4 \left(\Gamma ^2+\lambda ^2\right)+\Gamma _0^2\right)+1\right)\right) \\
   \fl
   \nonumber
  &+& 
  2  \lambda  Y \left(\alpha _{+0}+4 \lambda ^2 Y^2
   \alpha _{+0}-2 \lambda  Y \left(\alpha _{00}+2 \alpha _{++}\right)\right) \\
   \fl
   \nonumber
   &+&
   \alpha _{+-} \left(Y^4 \left(16 \left(\Gamma ^2+\lambda
   ^2\right)^2+\Gamma _0^4\right)+2 \Gamma _0^2 Y^2+1\right),\\
\fl \nonumber
\bar{A}_{++} &=&  \alpha _{++} \left(Y^4 \left(16 \left(\Gamma ^4-6 \Gamma ^2 \lambda ^2+\lambda ^4\right)+\Gamma _0^4\right)-6 \Gamma _0^2 Y^2+1\right) \\
\fl
\nonumber
&-&
2 Y \left(-\Gamma 
   \beta _{+0}+Y^3 \left(\Gamma _0 \left(\lambda  \left(-12 \Gamma ^2+\Gamma _0^2+4 \lambda ^2\right) \beta _{+0} \right. \right. \right. \\
   \fl
   \nonumber
   &+&
   \left.\left. \left.
   2 \Gamma _0 (\lambda -\Gamma ) (\Gamma
   +\lambda ) \left(\alpha _{00}-2 \alpha _{+-}\right)\right)+32 \Gamma  \lambda  \left(\Gamma ^2-\lambda ^2\right) \beta _{++}\right)  \right. \\
   \fl
   \nonumber
   &+&
   \left.
   Y^2
   \left(\Gamma  \left(-4 \Gamma ^2+3 \Gamma _0^2+12 \lambda ^2\right) \beta _{+0}+8 \Gamma  \Gamma _0 \lambda  \left(\alpha _{00}-2 \alpha
   _{+-}\right)-2 \Gamma _0^3 \beta _{++}\right) \right. \\
   \fl
   \nonumber
   &+&
   \left.
   \alpha _{+0} \left(-12 \Gamma  \Gamma _0 \lambda ^2 Y^3+\lambda  \left(1-3 \left(4 \Gamma
   ^2+\Gamma _0^2\right) Y^2\right) \right. \right.\\
   \fl
   \nonumber
   &+&
   \left. \left.
   \Gamma  \Gamma _0 Y \left(\left(4 \Gamma ^2+\Gamma _0^2\right) Y^2-3\right)+4 \lambda ^3 Y^2\right)-3 \Gamma _0 \lambda  Y
   \beta _{+0} \right. \\
   \fl
   \nonumber
   &-&
   \left.
   2 Y (\lambda -\Gamma ) (\Gamma +\lambda ) \left(\alpha _{00}-2 \alpha _{+-}\right)+2 \Gamma _0 \beta _{++}\right),\\
\fl \nonumber
\bar{A}_{0+} &=& 2 Y \left(2 \left(Y \left(\Gamma _0 \left(\lambda  \beta _{++} \left(4 Y^2 \left(\lambda ^2-3 \Gamma ^2\right)+3\right)-4 Y \left(\Gamma ^2+\lambda
   ^2\right) \beta _{+0}\right) \right. \right. \right. \\
   \fl
   \nonumber
   &+&
   \left.\left.\left.
   4 \left(\lambda ^2 \alpha _{+0}+\Gamma  \lambda  \left(\beta _{+0}-3 \lambda  Y \beta _{++}\right)+\Gamma ^3 Y \beta
   _{++}\right) \right. \right. \right. \\
   \fl
   \nonumber
   &+&
   \left.\left. \left.
   \Gamma  \Gamma _0^2 Y \left(-4 \Gamma  Y \alpha _{+0}+4 \lambda  Y \beta _{+0}+3 \beta _{++}\right)+\Gamma _0^3 \lambda 
   \left(-Y^2\right) \beta _{++}\right) \right. \right.\\
   \fl
   \nonumber
   &+&
   \left. \left.
   \alpha _{+-} \left(\lambda -\Gamma  \Gamma _0 Y\right) \left(Y^2 \left(4 \left(\Gamma ^2+\lambda
   ^2\right)+\Gamma _0^2\right)+1\right)-\Gamma  \beta _{++} \right. \right.\\
   \fl
   \nonumber
   &+&
   \left.\left.
   \alpha _{++} \left(-12 \Gamma  \Gamma _0 \lambda ^2 Y^3+\lambda  \left(3 \left(4
   \Gamma ^2+\Gamma _0^2\right) Y^2-1\right) \right. \right. \right. \\
   \fl
   \nonumber
   &+&
   \left. \left. \left.
   \Gamma  \Gamma _0 Y \left(\left(4 \Gamma ^2+\Gamma _0^2\right) Y^2-3\right)-4 \lambda ^3 Y^2\right)\right) \right. \\
   \fl
   \nonumber
   &+&
   \left.
   \alpha_{00} \left(\Gamma  \Gamma _0 Y-\lambda \right) \left(Y^2 \left(4 \left(\Gamma ^2+\lambda ^2\right)+\Gamma _0^2\right)+1\right)\right),\\
\fl \nonumber
\bar{A}_{+0} &=& 2 Y \left(-\Gamma _0 \beta _{+0}+Y^3 \left(4 \Gamma  \lambda  \left(4 \left(\Gamma ^2+\lambda ^2\right)+\Gamma _0^2\right) \beta _{+0} \right. \right. \\
\fl
\nonumber
&+&
\left. \left.
4 \Gamma _0
   \left(\Gamma ^3 \left(\alpha _{00}-2 \alpha _{+-}+2 \alpha _{++}\right)+\Gamma  \lambda ^2 \left(\alpha _{00}-2 \alpha _{+-}-6 \alpha
   _{++}\right)-6 \Gamma ^2 \lambda  \beta _{++}+2 \lambda ^3 \beta _{++}\right) \right. \right. \\
   \fl
   \nonumber
   &+&
   \left. \left.
   \Gamma _0^3 \left(\Gamma  \left(\alpha _{00}-2 \alpha
   _{+-}+2 \alpha _{++}\right)-2 \lambda  \beta _{++}\right)\right) \right.  \\
   \fl
   \nonumber
   &-&
   \left.
   Y^2 \left(4 \Gamma _0 \left(\Gamma ^2+\lambda ^2\right) \beta
   _{+0}+\Gamma _0^3 \beta _{+0}+4 \left(\Gamma ^2 \lambda  \left(\alpha _{00}-2 \left(\alpha _{+-}+3 \alpha _{++}\right)\right) \right.\right. \right.\\
   \fl
   \nonumber
   &+&
   \left. \left. \left.
   \lambda ^3
   \left(\alpha _{00}-2 \alpha _{+-}+2 \alpha _{++}\right)-2 \Gamma ^3 \beta _{++}+6 \Gamma  \lambda ^2 \beta _{++}\right) \right. \right. \\
   \fl
   \nonumber
   &+&
   \left. \left.
   \Gamma
   _0^2 \left(\lambda  \left(\alpha _{00}-2 \left(\alpha _{+-}+3 \alpha _{++}\right)\right)-6 \Gamma  \beta _{++}\right)\right)+4 \Gamma 
   \lambda  Y \beta _{+0} \right.\\
   \fl
   \nonumber
   &+&
   \left.
   \Gamma _0 Y \left(\Gamma  \left(\alpha _{00}-2 \alpha _{+-}-6 \alpha _{++}\right)+6 \lambda  \beta
   _{++}\right)-2 \Gamma  \beta _{++}\right)\\
   \fl
   \nonumber
   &+&
   \alpha _{+0} \left(Y^4 \left(16 \lambda ^4-\left(4 \Gamma ^2+\Gamma _0^2\right){}^2\right)+8 \lambda
   ^2 Y^2+1\right)-2 \lambda  Y \left(\alpha _{00}-2 \alpha _{+-}+2 \alpha _{++}\right),\\
\fl \nonumber
\bar{B}_{+} &=& \beta _+ \left(Y^2 \left(-4 \Gamma ^2-\Gamma _0^2+4 \lambda ^2\right)+1\right)-2 \lambda  Y \left(\beta _0+4 \alpha _+ \Gamma  Y\right)+2 \Gamma _0 Y
   \left(\alpha _++\beta _0 \Gamma  Y\right),\\
\fl \nonumber
\bar{B}_{++} &=&  2 Y \left(\Gamma _0 \left(Y \left(3 \Gamma  \beta _{+0}+\lambda  \alpha _{+0} \left(-4 Y^2 \left(\lambda ^2-3 \Gamma ^2\right)-3\right)+4 \Gamma  Y^2
   \left(\Gamma ^2-3 \lambda ^2\right) \beta _{+0} \right.\right. \right.\\
   \fl
   \nonumber
   &+&
   \left. \left. \left.
   4 Y (\lambda -\Gamma ) (\Gamma +\lambda ) \left(\alpha _{00}-2 \alpha _{+-}\right)\right)+2 \alpha
   _{++}\right) \right.\\
   \fl
   \nonumber
   &+&
   \left.
   \Gamma  \left(4 \lambda  Y \left(\alpha _{00}+Y \left(8 Y (\lambda -\Gamma ) (\Gamma +\lambda ) \alpha _{++}-3 \Gamma  \beta
   _{+0}\right)-2 \alpha _{+-}\right) \right.\right. \\
   \fl
   \nonumber
   &+&
   \left.\left.
   \alpha _{+0} \left(4 Y^2 \left(\Gamma ^2-3 \lambda ^2\right)-1\right)\right)+\Gamma _0^2 Y^2 \left(3 \lambda 
   \beta _{+0}+\Gamma  \left(3 \alpha _{+0}-4 \lambda  Y \left(\alpha _{00}-2 \alpha _{+-}\right)\right)\right) \right. \\
   \fl
   \nonumber
   &+&
   \left.
   \Gamma _0^3 Y^2 \left(\lambda  Y
   \alpha _{+0}-\Gamma  Y \beta _{+0}-2 \alpha _{++}\right)\right)+8 \lambda ^3 Y^3 \beta _{+0}-2 \lambda  Y \beta _{+0}\\
   \fl
   \nonumber
   &+&
   \beta _{++} \left(Y^4
   \left(\Gamma _0^4-16 \left(\Gamma ^4-6 \Gamma ^2 \lambda ^2+\lambda ^4\right)\right)-6 \Gamma _0^2 Y^2+1\right),\\
\fl \nonumber
\bar{B}_{0+} &=& -2 Y \left(-2 \left(4 Y \left(\Gamma +\Gamma _0 \lambda  Y\right) \left(\alpha _{+0} \left(\lambda +\Gamma  \Gamma _0 Y\right)+\beta _{+0} \left(\Gamma
   -\Gamma _0 \lambda  Y\right)\right) \right. \right. \\
   \fl
   \nonumber
   &+&
   \left. \left.
   \alpha _{+-} \left(\Gamma +\Gamma _0 \lambda  Y\right) \left(Y^2 \left(4 \left(\Gamma ^2+\lambda ^2\right)+\Gamma
   _0^2\right)+1\right) \right. \right. \\
   \fl
   \nonumber
   &+&
   \left. \left. 
   \alpha _{++} \left(-12 \Gamma ^2 \Gamma _0 \lambda  Y^3+4 \Gamma ^3 Y^2+\Gamma _0 \lambda  Y \left(Y^2 \left(\Gamma _0^2+4
   \lambda ^2\right)-3\right) \right. \right. \right. \\
   \fl
   \nonumber
   &+&
   \left. \left. \left.
   \Gamma  \left(1-3 Y^2 \left(\Gamma _0^2+4 \lambda ^2\right)\right)\right)\right)+\alpha _{00} \left(\Gamma +\Gamma _0 \lambda 
   Y\right) \left(Y^2 \left(4 \left(\Gamma ^2+\lambda ^2\right)+\Gamma _0^2\right)+1\right) \right. \\
   \fl
   \nonumber
   &+&
   \left.
   \beta _{++} \left(-24 \Gamma  \Gamma _0 \lambda ^2
   Y^3+\lambda  \left(6 \left(4 \Gamma ^2-\Gamma _0^2\right) Y^2+2\right)\right.\right. \\
   \fl
   \nonumber
   &+&
   \left.\left.
   2 \Gamma  \Gamma _0 Y \left(\left(4 \Gamma ^2-\Gamma _0^2\right) Y^2+3\right)-8
   \lambda ^3 Y^2\right)\right),\\
\fl \nonumber
\bar{B}_{+0} &=& 2 Y \left(\Gamma _0^3 Y^2 \left(\alpha _{+0}+Y \left(-\lambda  \alpha _{00}+2 \lambda  \left(\alpha _{+-}+\alpha _{++}\right)+
2 \Gamma  \beta _{++}\right)\right) \right.\\
\fl
\nonumber
&+&
\left.
\Gamma _0^2 Y^2 \left(6 \lambda  \beta _{++}-\Gamma  \left(\alpha _{00}
-4 \lambda  Y \alpha _{+0}-2 \alpha _{+-}+6   \alpha _{++}\right)\right) \right. \\
\fl
\nonumber
&+&
\left.
4 \Gamma  Y \left(\lambda  \left(\alpha _{+0}+4 \lambda ^2 Y^2 \alpha _{+0}-\lambda  Y \left(\alpha _{00}-2 \alpha
   _{+-}+6 \alpha _{++}\right)\right) \right. \right. \\
   \fl
   \nonumber
   &+&
   \left. \left.
   \Gamma ^2 Y \left(2 \left(2 \lambda  Y \alpha _{+0}+\alpha _{+-}+\alpha _{++}\right)-\alpha_{00}\right)\right)+
   \Gamma _0 \left(\alpha _{+0} \left(4 Y^2 \left(\Gamma ^2+\lambda ^2\right)+1\right) \right. \right. \\
   \fl
   \nonumber
   &+&
   \left. \left.
   4 Y^3 \left(-\Gamma ^2 \lambda  \left(\alpha
   _{00}-2 \alpha _{+-}+6 \alpha _{++}\right)+\lambda ^3 \left(2 \left(\alpha _{+-}+\alpha _{++}\right)-\alpha _{00}\right)-2
   \Gamma ^3 \beta _{++}+6 \Gamma  \lambda ^2 \beta _{++}\right) \right. \right. \\
   \fl
   \nonumber
   &-&
   \left.\left. 
   Y \left(\lambda  \left(\alpha _{00}-2 \alpha _{+-}+6 \alpha
   _{++}\right)+6 \Gamma  \beta _{++}\right)\right)+\Gamma  \left(2 \left(\alpha _{+-}+\alpha _{++}\right)-\alpha _{00}\right) \right. \\
   \fl
   \nonumber
   &+&
   \left.
   2 \lambda  \beta _{++} \left(4 Y^2 \left(\lambda ^2-3 \Gamma ^2\right)-1\right)\right)+\beta _{+0} \left(Y^4 \left(-\left(\left(\Gamma _0^2+4 \lambda
   ^2\right){}^2-16 \Gamma ^4\right)\right)+8 \Gamma ^2 Y^2+1\right),
\end{eqnarray}

\ni all subject to the constraining relations which will be determined in the following subsections.

\subsubsection{\bf{The subclass of linear Hamiltonians}}\label{sec_quad}

\ \\

The system of nonlinear equations (\ref{setconst}) is not of trivial solution and we shall start by analysing the behaviour of the linear $su(2)$ Hamiltonian according to this framework. For that matter, we first consider the subclass of (\ref{constraints}) with $\alpha_{++}=\beta_{++}=\alpha_{+0}=\beta_{+0}=0$, such that (\ref{llgeneral}) considerably simplifies to:
\begin{equation}\label{llinear}
\fl
H= {i} \beta_0 L_0 + (\alpha_+ + {i} \beta_+)L_+ + (\alpha_+ - {i} \beta_+)L_- + \alpha_{00} L_0^2 + \alpha_{+-}(L_+L_- + L_- L_+)
\end{equation}
The non-Hermiticity of the operator above comes solely from the first term, all others, as a group, being invariant under Hermitian conjugation.
A metric can be found by solving the constraining equations (\ref{constraints}) as long as $\alpha_{+-} = \frac{1}{2}\alpha_{00}$, meaning the above becomes just the linear Hamiltonian shifted by the Casimir operator $L^2$. The Dyson operator, and consequently the metric, is then characterised by
the parameters $\lambda, \Gamma, \Gamma_0$ such that the combination $\theta^2 =  4(\lambda^2 + \Gamma^2) -\Gamma_0^2$ obeys
\begin{eqnarray}\label{simthetalin}
\fl
\frac{\tanh (\theta )}{\theta }=\frac{\beta _0}{\Delta -4 \left(\alpha _+ \Gamma -\beta _+ \lambda \right)} \\
\fl
\nonumber
\Delta = \pm  \sqrt{16( \beta_+\lambda  -\alpha_+ \Gamma)^2+8\beta_0\Gamma_0( \alpha_+\lambda+\beta_+\Gamma)-\beta_0^2(\Gamma_0^2+4(\lambda^2+\Gamma^2))}.
\end{eqnarray}

The above constraints is to be used in (\ref{transllgeneral}) when calculating the Hermitian isospectral partner of (\ref{llinear}), together with the specifications of the linear Hamiltonian, namely $\alpha_{++}=\beta_{++}=\alpha_{+0}=\beta_{+0}=0$.
A direct comparison between the non-Hermitian Hamiltonian and its Hermitian counterpart is presented in Figure 1, where we have specified all variables but one. The initial parameters in (\ref{llinear}) are $\alpha_+$, $\alpha_-$ and $\beta_0$. By fixing the first two of them and reparameterizing the last one in terms of $\theta$, one can obtain a non-Hermitian Hamiltonian in terms of this latter variable. This choice is such that the free parameter measures the non-Hermiticity coming from the $L_0$ term. Using a finite dimensional matrix representation of the generators of the type
\begin{equation}\label{spinbasis}
\fl \qquad
L_\pm | l, m \rangle = \sqrt{(l \mp m)(l \pm m +1)}| l, m \pm 1 \rangle , \qquad
L_0 | l, m \rangle =m  | l, m  \rangle,
\end{equation}
the diagonalisation of the Hamiltonian is possible and the plot of the eigenvalues as a function of $\theta$, is presented on the left. Here, it was used a representation of dimension $11$ associated to $l=5$ and $m=-5,-4,...,0,...,4,5$. For the right hand side, the spectrum is obtained by diagonlizing the transformed Hamiltonian  (\ref{transllgeneral}), with $\alpha_{++}=\beta_{++}=\alpha_{+0}=\beta_{+0}=0$ and subject to (\ref{simthetalin}). The diagonalization of this Hermitian counterpart involves other 3 free parameters, $\lambda, \Gamma, \Gamma_0$, and $\theta$ has an interpretation in terms of the parameters of the metric, so that the constraint (\ref{simthetalin}) is always satisfied for $\theta^2 \leq 8$. As one can see, the spectral curves for arbitrary values of the metric parameters, coincide, as expected, reinforcing the results presented.

\begin{figure}[h!]
\begin{center}
\includegraphics[width=0.45\textwidth]{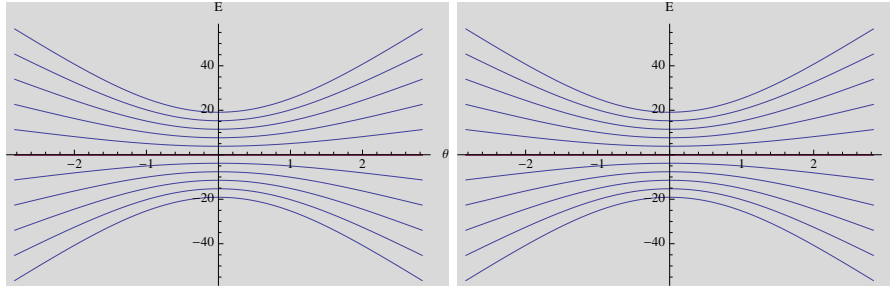}
\includegraphics[width=0.45\textwidth]{cptspect.eps}
\fl
\caption{On the left, the spectrum of the linear non-Hermitian $su(2)$ Hamiltonian (\ref{llinear}) with $\alpha_+=4, \beta_+=4, \alpha_{00}=\alpha_{+-}=0$ held fixed and $\beta_0=\frac{64 \sqrt{8-\theta^2} \sinh ^2(\theta)}{\theta^2+16 \sinh ^2(\theta)}$, which is the term responsible for the breaking of Hermiticity, being parameterized in terms of $\theta$. An independent diagonalization is shown on the right, where we have the spectrum of (\ref{transllgeneral}), with $\alpha_{++}=\beta_{++}=\alpha_{+0}=\beta_{+0}=0$ and subject to (\ref{simthetalin}), again as a function of $\theta$ in the interval $[-2\sqrt{2},2\sqrt{2}]$, with $\alpha_+=4, \beta_+=4, \alpha_{00}=\alpha_{+-}=0, \beta_0=\frac{64 \sqrt{8-\theta^2} \sinh ^2(\theta)}{\theta^2+16 \sinh ^2(\theta)}$ besides $\lambda=1, \Gamma=1, \Gamma_0=\sqrt{8-\theta^2}$. The graphs coincide, as expected, since the Hamiltonians are related by a similarity transformation characterized by $\lambda, \Gamma, \Gamma_0$.}
\label{lineargraph}
\end{center}
\end{figure}

The subcase of (\ref{llinear}) where $\alpha_{00}=\alpha_{+-}=0$ corresponds to the non interactive limit of the Bose-Hubbard model when $\beta_+ =0$. It is also the $su(2)$ analogue of the so-called Swanson Hamiltonian \cite{swanson}, defined for $su(1,1)$
\begin{eqnarray}
H_S = \mu _{0}K_{0}+\mu _{-}K_{-}+\mu _{+}K_{+}
\end{eqnarray}
\ni with energy spectrum
$E_{n}^{S}= \left( n+ \frac{1}{2} \right) \sqrt{\frac{\mu_{0}^{2}}{4}-\mu_{-}\mu_{+}}$, in agreement with (\ref{twoboson}). Here $\mu_{0,\pm}$ are nothing but parameters specifying an   extension of the harmonic oscillator with characteristic frequency $\mu_0$, which can be recovered when $\mu_+ = \mu_-=0$; otherwise, when $\mu_+ \neq \mu_-$, non-Hermiticity arises.
In a similar way as what was done for the Swanson model, one can express the Hamiltonian as an isospectral deformation of a multiple of the operator $K_0$, or $L_0$ in this situation. The determination of this proportionality factor, similarly to equation (\ref{HtoL0}), allows us to find the eigenvalues in the present case to be
\begin{equation}\label{eigenlin}
E_{n}= n \sqrt{(\alpha_+^2+\beta_+^2)-\frac{\beta_{0}^{2}}{4}},
\end{equation}
in accordance with the result of the previous section and describing the behaviour of the spectra shown in Figure 1. The real energy spectra of these Hamiltonians are complementary: whereas $E_{n}^{S} \in \mathbb{R}$ for $\mu _{0}^{2}>4\mu _{+}\mu _{-}$ one has $E_{n}\in \mathbb{R}$ for $\tilde{\mu}_{0}^{2} \equiv \beta_0^2 <4\tilde{\mu}_{+}\tilde{\mu}_{-} \equiv 4(\alpha_+^2+\beta_+^2)$ for the linear $su(2)$ Hamiltonian. Alternatively to its direct derivation, the equation above could also be inferred by the correspondence between the generators $L$ and $K$, mentioned in the end of subsection \ref{subframe}.

The Swanson linear Hamiltonian, however, is somewhat trivial and the quadratic terms are the ones giving rise to more interesting behaviour, such as interaction. Therefore the understanding of quadratic problems is relevant and shall be pursued in the next subsection.

\subsubsection{\bf{The general quadratic Hamiltonian}}\label{sec_quad}

\ \\


Having presented the general framework allowing us to construct a consistent description of non-Hermitian Hamiltonians and shown the behaviour of the simpler linear subclass of Hamiltonians, we can now tackle the more involved situation where quadratic terms are added.
To find the constraints for more general quadratic Hamiltonians  (\ref{llgeneral}) to be mapped into a Hermitian counterpart by a Dyson operator of the form (\ref{bigeta}) we again apply the ansatz to the Hamiltonian and demand the result to be Hermitian.

It is important to note, that the $ \eta$ operator preserves the powers of the angular momentum components appearing in the Hamiltonian. That is, it maps the linear part to a linear counterpart, and the quadratic part to a purely quadratic part, without linear terms.
The quadratic part adds two additional constraints, besides those involving the coefficients of the terms linear $L_0=L_z$ in the Hamiltonian.
Nevertheless, the fact that the Hermitian conjugate of $L_0 L_\pm$ is $L_\mp L_0$ implies that in order to rewrite the latter in terms of the former, the commutation relations need to used, introducing the non-Hermiticity of the quadratic parts in the linear ones.
As a consequence the constraint obtained in the linear case gets extended, with the quadratic part in general leading to additional condition,  in the quadratic case but stays valid in the limit as $\alpha_{ij} \ra 0$ and $\beta_{ij} \ra 0$.
Thus, although the introduction of the quadratic parameters increases the dimension of the manifold defined by the constraints, the range of linear coefficients for which a linear $ \eta$ operator can be found, does not increase due to an additional quadratic terms, and lies within the region $\varsigma_x^2+\varsigma_y^2-\varsigma_z^2=4(\alpha_+^2+\beta_+^2)-\beta_{0}^{2}>0$.


However, the class of $ \eta$ operators (\ref{bigeta}) is larger than necessary. For any Hamiltonian that is mapped to a Hermitian counterpart by an $ \eta$ operator of the form (\ref{eqn_eta_ansatz}) with $\Gamma_0 =\delta\neq0$ there exists another $ \eta$ with $\Gamma_0 = \delta=0$ that also maps the non-Hermitian to a Hermitian Hamiltonian. This is because the $L_0$ contribution for this particular kind of ansatz can be factored out, in a similar fashion as the Gauss decomposition of (\ref{bigeta}) leading to (\ref{gaussdec}), as a unitary operator, which cannot transform a non-Hermitian operator to a Hermitian one and vice-versa.  Thus, to investigate the quadratic non-Hermitian Hamiltonian, we will use the simpler ansatz
\begin{equation}\label{tildeta}
\tilde{\eta}=\exp{\left[2\left\{(\lambda + {i}\Gamma)L_+ + (\lambda - {i} \Gamma)L_- \right\}\right]}, \label{eqn_eta_ansatz}.
\end{equation}
\ni where the $L_0$ part has been removed.
Further, it is useful to consider the parameters in the metric to be proportional, $\Gamma=\lambda \nu$. The exponent in the metric then reads
$T=2\lambda[(1+i \nu)L_+ + (1-i \nu)L_-]$. This $ \eta$ operator can be interpreted as a rotation of the space $x,y,z$ around an axis in the $x,y$ plane by an imaginary angle, that is a boost.

Here we present results for the Hermitian metric, with $\Gamma_0=0$, but analogous expressions for $\Gamma_0 \neq 0$ can clearly also be found, as argued before. We present results for the latter in an appendix section. By solving the constraints (\ref{constraints})  coming from the Hermiticity conditions on the transformed Hamiltonian (\ref{dysonherm}), we end up with relations fixing the parameters in the metric, $\lambda$ and $\\nu$, through the quantity $\theta$ and also restraining some of the coefficients in the Hamiltonian.
The metric transformations which map the non-Hermitian Hamiltonians into Hermitian counterparts must satisfy the following relation
\begin{eqnarray}\label{genconst}
\rm tanh (\theta) = \frac{2\beta_+ - \beta_{+0} -(2\alpha_+ -
\alpha_{+0})\nu \pm \bar{\Delta}}{\beta_0 \sqrt{1+\nu^2}} \\
\nonumber
\bar{\Delta} = \sqrt{(2\beta_+ - \beta_{+0} -(2\alpha_+ -
\alpha_{+0})\nu)^2-\beta_0^2(1+\nu^2)}
\end{eqnarray}

\ni with $\theta = 2\lambda\sqrt{\nu^2+1}$ and further constraints
\begin{eqnarray}\label{quadaconst}
\fl
\nonumber &\alpha_{00} & = 2\alpha_{+-}+\frac{(2\alpha_+ -
\alpha_{+0})(\alpha_{+0}-2\beta_{+0}(\nu-1))}{2\beta_0(\nu-1)}+ \\
\fl
\nonumber
&+& \frac{\beta_0(\alpha_{+0}+\beta_{+0}\nu)}{2\beta_+
- \beta_{+0}+(\alpha_{+0}-2\alpha_+)\nu}+2\alpha_{++}\frac{\nu^2+1}{\nu^2-1}+ \\
\fl
\nonumber &+& \frac{\alpha_{+0}(\alpha_{+0}-4(\beta_+ -
\beta_{+0}))-(\alpha_{+0}^2+2\beta_{+0}(\beta_{+0}-2\beta_+))\nu -
2\alpha_+(\alpha_{+0}+2\beta_{+0}-\alpha_{+0}\nu)}{2\beta_0(\nu^2-1)} \\
\fl
&\beta_{++} & = \frac{(\beta_{+0}-\alpha_{+0}\nu)(\beta_{+0}-2\beta_{+} + (2\alpha_+ - \alpha_{+0})\nu) - 4 \alpha_{++} \beta_{0} \nu}{2\beta_0(\nu^2-1)}.
\end{eqnarray}

Setting $\nu=0$ in the expressions above, the constraints in the metric parameters and Hamiltonian coefficients simplify significantly, constituting simpler solutions.
The linear limit ($\alpha_{ij}=\beta_{ij}=0$) reproduces as expected the results in (\ref{simthetalin}). In these cases, for instance, the constraining equations might be more conveniently expressed in terms of $\tanh(2\theta)$. If this is done, care must be taken in order to make sure no spurious solutions are added as there is an ambivalence between $\tanh(\theta)$ and $\coth(\theta)$ in this situation: $\tanh(2\theta) = \frac{2\tanh{\theta}}{1+\tanh^2{\theta}} = \frac{2\coth{\theta}}{1+\coth^2{\theta}}$. 

The possibility of finding exact conditions for the so-called pseudo-quasi-Hermiticity to be present in a system is without a doubt a strong point of the current method. Moreover, this approach gives at the same time, not only information about the metric with respect to which the Hamiltonian has a unitary evolution with real eigenvalues, but also allows us to construct Hermitian isospectral partners. In order to do so, one just has to act with the Dyson similarity transformation on the original Hamiltonian, as described by {dysonherm}. The Hermitian counterpart for (\ref{llgeneral}) constructed in this way is specified by (\ref{transllgeneral}), with $\Gamma_0 =0, \Gamma = \nu \lambda$, and constrained according to (\ref{quadaconst}).

In Figure 2 below we confirm again that the original non-Hermitian quadratic $su(2)$ Hamiltonian has a spectrum which exactly matches that of the constructed Hermitian counterpart with $\tilde{\eta}$ (\ref{tildeta}), providing another consistency check. Similarly to the analysis of Figure 1, again we fix the parameters in order to satisfy the Hermiticity constraints (\ref{genconst}) and leave one of them, $\beta_0$, free to vary. The diagonalization is done with the angular momentum representation for $l=5$. However, now the intervals in both graphs are chosen to be different. Whereas the first graph gives a more general picture of the behaviour of the eigenvalues, the interval in the subsequent graph is also interesting as it shows in more detail that the apparent level-crossing of the characteristic values observed on the left actually corresponds to avoided crossing of levels, as shown on the right. Due to the presence of quadratic interacting terms, the structure of the energy levels is considerably richer than that of the linear case, in Figure 1, deviating from the simple harmonic oscillator. Differently from the linear case, now the Hamiltonian cannot be put in a form proportional to $L_0$ only, so the analogue of equation (\ref{eigenlin}) cannot be presented here.

\begin{figure}[h!]
\begin{center}
\includegraphics[width=0.45\textwidth]{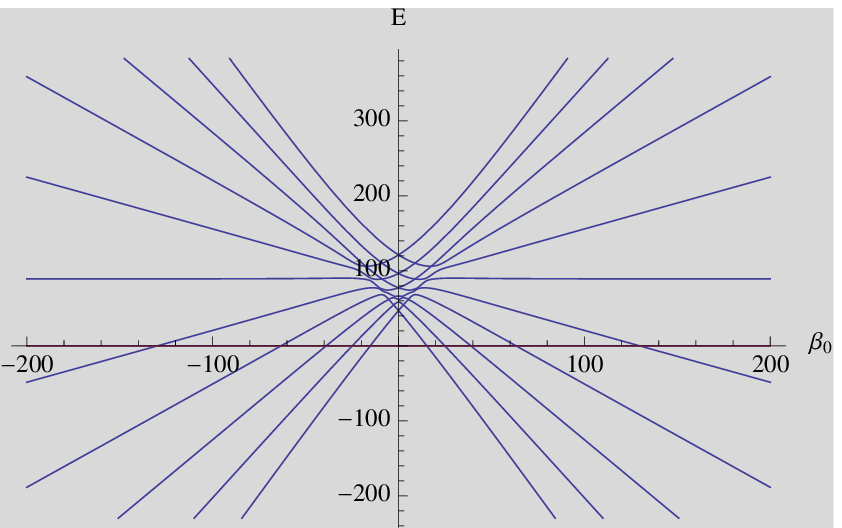}
\includegraphics[width=0.45\textwidth]{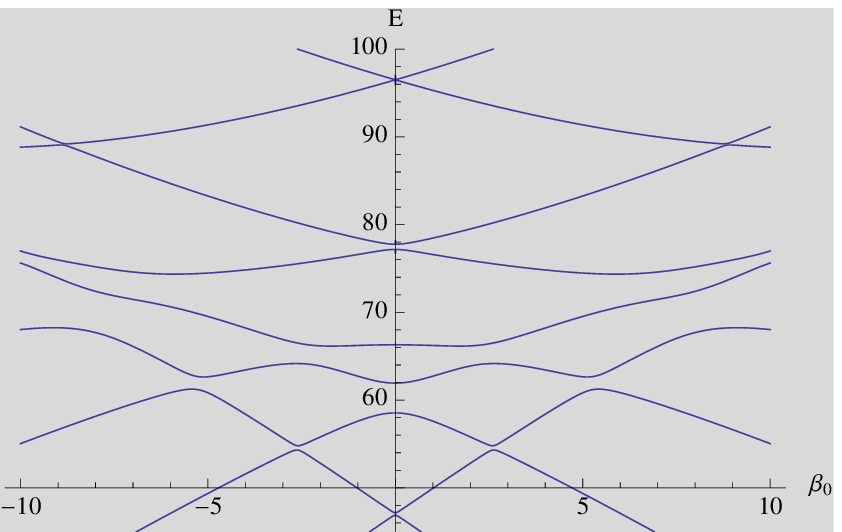}
\fl
\caption{On the left, the spectrum of the linear non-Hermitian $su(2)$ Hamiltonian (\ref{llgeneral}) with $\alpha_{+-}=1, ~\alpha_{++}=1, ~\alpha_{+0}=2,~ \beta_{+0}=1, ~\alpha_+=1$ held fixed and
$\beta_+=\frac{1}{4} \left[2 (2 \alpha_+ \nu -\alpha_{+0} \nu +\beta_{+0})+\beta_0 \sqrt{\nu ^2+1} \left(\tanh \left(2 \lambda  \sqrt{\nu
   ^2+1}\right)+\coth \left(2 \lambda  \sqrt{\nu ^2+1}\right)\right)\right]$, which is the term responsible for the breaking of Hermiticity being parameterized in terms of $\theta$. Independently, the diagonalization shown on the right depicts the spectrum of (\ref{transllgeneral}), again as a function of $\theta$ in the interval $[-2\sqrt{2},2\sqrt{2}]$, with $\alpha_{+-}=1, ~\alpha_{++}=1, ~\alpha_{+0}=2,~ \beta_{+0}=1, ~\alpha_+=1$ held fixed and
$\beta_+=\frac{1}{4} \left[2 (2 \alpha_+ \nu -\alpha_{+0} \nu +\beta_{+0})+\beta_0 \sqrt{\nu ^2+1} \left(\tanh \left(2 \lambda  \sqrt{\nu
   ^2+1}\right)+\coth \left(2 \lambda  \sqrt{\nu ^2+1}\right)\right)\right]$ together with the parameters of the metric $\tilde{\eta}$ to be $\lambda=0.3, ~\nu=0.2$. Once more, the graphs coincide, a result that serves as a check that the similarity transformation constructed is correct. The different intervals used for the graphs shown give a general behaviour of the spectum, on the left, whereas on the right one has more details, particularly regarding the avoided level-crossing, which is unnoticed in the picture on the left.}
\label{lineargraph}
\end{center}
\end{figure}

Whereas in the linear case the class of pseudo-quasi-Hermitian operators is identical to the class of operators for which an $ \eta$ operator of the form (\ref{bigeta}) exists, and the latter can always be found via direct diagonalisation, these classes of Hamiltonians do not coincide in the quadratic case. Only a subclass of Hamiltonians quadratic in $su(2)$ can be diagonalised by a similarity transformation with linear exponents in the generators of $su(2)$. In addition there are non-Hermitian Hamiltonians of the type (\ref{llgeneral}) for which such $ \eta$ operators can be found, that map them to Hermitian, but not diagonal counterparts. A whole class of such Hamiltonians with a nontrivial spectrum can for example be obtained from the linear system, by adding a Hermitian quadratic part that commutes with the metric obtained for the linear part.

It is also interesting to note that by fixing the parameters in the Hamiltonian arbitrarily and letting one of them, $\beta_0$ again, vary freely, without respect to the constraints (\ref{genconst}) we observe in Figure 3 the emergence of complex eigenvalues for the problem, a behaviour very similar to that described in \cite{ptbec}. This is a manifestation of the broken \pt-symmetry, in which case despite having a \pt-symmetric Hamiltonian as we do, the eigenfunctions of the corresponding states at not invariant under parity and time reversal.
For the values chosen a metric of the form (\ref{metansatz})  can only be found for either $\beta_0 =0$ or $\beta_0 = 2.4733$ but the graph shows us that the region of unbroken \pt-symmetry is larger and we expect that a more general metric should exist to cover the whole sector.

\begin{figure}[h!]
\begin{center}
\includegraphics[width=0.45\textwidth]{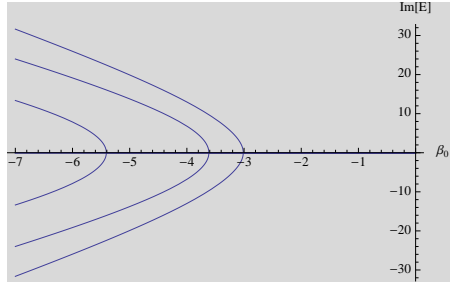}
\fl
\caption{Imaginary part of the spectrum of quadratic non-Hermitian $su(2)$ Hamiltonian (\ref{llgeneral}) as a function of $\beta_0$ with
 $\alpha_{+-}=1, ~\alpha_{++}=1, ~\alpha_{+0}=2,~ \beta_{+0}=1,~ \alpha_+=1, ~\beta_+=2$. An equivalent spectrum for the Hermitian counterpart (\ref{transllgeneral})  can be obtained, for instance, with $\lambda=0.3, ~\nu=0.2$ in $\tilde{\eta}$. Due to the symmetry of the spectrum with respect to $\beta_0$, only a negative interval is shown. The energy levels cease to be real approximately when $|\beta_0|>3$.}
\label{lineargraph}
\end{center}
\end{figure}

Considering the vanishing of the quadratic parameters $\alpha_{ij}, \beta_{ij} =0$,  condtions (\ref{quadaconst}) are automatically satisfied whereas (\ref{genconst}) can be rewritten as
\begin{equation}
\tanh \left(4 \sqrt{\lambda ^2+\Gamma ^2}\right)=\frac{\beta_0 \sqrt{\lambda ^2+\Gamma ^2}}{2 (\beta_+ \lambda
   -\alpha_+ \Gamma )},
\end{equation}
which coincides with (\ref{simthetalin}), when $\Gamma_0=0, \Gamma=\nu \lambda$, written in terms of the double angle. Parametrising $\lambda = \frac{1}{4}\phi \cos\vartheta, \Gamma = \frac{1}{4}\phi \sin\vartheta$, the above equation can be  interpreted as the orthogonality condition between the vectors $(2\alpha_+ , -2\beta_+ , \beta_0)$ and $(\sin\vartheta ~\sinh \phi, \cos\vartheta ~\sinh \phi, \cosh \phi)$ \cite{evaprivate}.
The second vector points from the origin to a hyperboloid of two sheets, and can be written as  $(x, y, \pm \sqrt{1+x^2+y^2})$ for any $x,y \in \mathds{R}$. Thus, the absolute value of the slope of these vectors in the direction of $z$ is always larger than $1$. That is, if and only if the slope of $(2\alpha_+ , -2\beta_+ , \beta_0)$ in $z$-direction lies between $-1$ and $+1$ can it be orthogonal to a vector of this form. This gives the constraint  $|\beta_0|\leq2\sqrt{\alpha_+^2+\beta_+^2}$,
and it is now straightforwardly verified that the $ \eta$ operator which was extracted from the diagonalisation fulfils this constraint, in agreement with
(\ref{eigenlin}).  Due to the existence of more parameters in the general quadratic Hamiltonian, a simple geometrical interpretation of this form cannot be achieved.

Now that a systematic study of the metric operators associated to the problem (\ref{llgeneral}) has been achieved, the constraints specified by (\ref{genconst}) and (\ref{quadaconst}) can be used to investigate the interacting Bose-Einstein Hamiltonian (\ref{hameva}), presented as a motivating example. This is simply accomplished by taking $\beta_+ = \beta_{++} = \beta_{+0} =\alpha_{+0}=\alpha_{++}= \alpha_{+-} =0$ in (\ref{equivaa}). As a result, the corresponding system (\ref{setconst}) does not have a consistent solution. This problem of not being able to find a metric of the proposed form motivates us to find alternative Dyson operators.

Next, we shall discuss  the existence of more general Dyson operators, by investigating the possibility of $ \eta$ operators with an exponent including quadratic terms instead of only linear operators.

\subsection{Quadratic metrics}

There are examples found in \cite{afkk}  where one is faced with non-Hermitian Hamiltonians having real spectra but for which a metric of the form (\ref{metansatz}) could not be found. This was only a confirmation of the fact that the ansatz used is far from being the most general.
A priori there is no reason why a non-Hermitian quadratic $su(2)$ Hamiltonian should be mapped to a Hermitian counterpart by an $\eta$ operator of the simple linear form. From the known solvable examples of Heisenberg-Weyl type, it is for example well known, that the metric often is not an analytic function of the coordinate and momentum operators and might even involve nonlocal expressions. It is expected that the situation for $su(2)$ Hamiltonians is in general not different. Thus, the class of non-Hermitian Hamiltonians of the type (\ref{llgeneral}) with real eigenvalues is expected to be larger than the class fulfilling the constraints in the last section  that is brought to a Hermitian counterpart by a simple $\eta$ with a linear exponent.


However, finding more general metrics soon goes beyond the capability of the approach followed up in the present paper, as we were using an explicit ansatz for the metric together with the Lie algebraic properties special to this ansatz.
With that in mind we can choose an operator whose exponent is, instead, purely quadratic in the operators  $L_0, L_\pm$,
\begin{equation}\label{quadmet}
\eta = e^{\zeta_{-}L_-^2}e^{\zeta_{0}L_0^2}e^{\zeta_{+}L_+^2}.
\end{equation}

For that purpose we make use of the following actions
\begin{eqnarray}\label{eqn_action_quad}
\nonumber e^{\zeta_{\pm}L_\pm^2} L_0 e^{-\zeta_{\pm}L_\pm^2} & = & L_0 \mp 2\zeta_\pm L_\pm^2, \\
\nonumber e^{\zeta_{+}L_+^2} L_- e^{-\zeta_{+}L_+^2}  & = & L_- + 2\zeta_+ (L_+ + 2  L_+ L_0 ) - 4\zeta_{+}^2  L_+^3,\\
\nonumber e^{\zeta_{-}L_-^2} L_+ e^{-\zeta_{-}L_-^2}  & = & L_+ - 2\zeta_- (L_- + 2  L_0 L_- ) - 4\zeta_{-}^2  L_-^3,\\
 e^{\zeta_{0}L_0^2} L_+ e^{-\zeta_{\pm}L_0^2} & = & e^{+\zeta_0} \left( L_+  e^{+2\zeta_0 L_0}\right), \\
\nonumber e^{\zeta_{0}L_0^2} L_- e^{-\zeta_{\pm}L_0^2} & = & e^{-\zeta_0} \left( e^{-2 \zeta_0 L_0 } L_- \right),
\end{eqnarray}

The expressions above lead to
\begin{eqnarray}\label{quadmetaction}
\nonumber \eta \, L_0 \, \eta^{-1} &=& c_0 - 2\zeta_+ e^{2\zeta_0}\left( c_1 e^{2\zeta_0 c_0} \right)^2, \\
\eta \, L_+ \, \eta^{-1} &=& e^{\zeta_0} \left( c_1 e^{2\zeta_0 c_0} \right), \\
\nonumber \eta \, L_- \, \eta^{-1} &=& e^{-\zeta_0}e^{-2\zeta_0 c_0}L_- +2\zeta_+e^{\zeta_0}\left( c_1 e^{2\zeta_0 c_0} \right)+ \\
\nonumber &&+4\zeta_+ e^{3\zeta_0}\left( c_1 e^{2\zeta_0 c_0} \right)c_0 -
4\zeta_+^2 e^{\zeta_0} \left( c_1 e^{2\zeta_0 c_0} \right)^3,
\end{eqnarray}

\noindent where we have abbreviated
\begin{eqnarray}
c_0 & \equiv & L_0 + 2\zeta_- L_-^2, \\
\nonumber c_1 & \equiv & L_+ -2\zeta_- L_- -4 \zeta_- L_0 L_- -4 \zeta_-^2 L_-^3.
\end{eqnarray}

From this actions we can already extract a number of interesting observations. First of all, we observe that the first equation in (\ref{eqn_action_quad}), when read from right to left so to speak, is already an example of a non-Hermitian Hamiltonian that is mapped to a Hermitian counterpart by a metric with a quadratic exponent. This is a trivial example, because in the standard basis of angular momentum it is immediately obvious that the eigenvalues of the non-Hermitian operator $ L_0+2 \zeta_{\pm}  L_{\pm}^2$ are identical to those of $ L_0$, and thus real, as in this basis, it corresponds to an upper/lower triangular matrix with the eigenvalues of $L_0$ on the diagonal. This example, is not of the type (\ref{llgeneral}) we have been investigating in the present paper, since there we cannot vanish $L_+$ or $L_-$ independently. However, we can easily construct an example of this type, by a rotation, so that the Hamiltonian on the right side of the first equation in (\ref{eqn_action_quad}) is in fact of the type (\ref{llgeneral}).

Secondly, due to the fact that an exponential of a product of $su(2)$ operators does not preserve the power of the generators $L$ upon which they act adjointly, the mapping above is clearly not a linear transformation. This strongly complicates the structure of constraints obtained from demanding the transformed Hamiltonian be Hermitian. In order to overcome this difficulty in formulating condition (\ref{dysonherm}), another possibility is to consider for simplicity a matrix representation for the problem in an attempt to construct a metric with quadratic exponents, as in (\ref{quadmet}). When calculating it, one can make use of the fact that, unlike $su(1,1)$, the algebra $su(2)$ admits finite dimensional representation in order to analyse the system more easily by working with a specific realisation. The simplest of these is the two-dimensional spin $\frac{1}{2}$ representation, generated by Pauli $\sigma$-matrices. However, the simplicity of such representation does not always play in our favour. If we make use of well known properties
\begin{equation}
\sigma_i \cdot \sigma_j = \delta_{ij}  \mathbb{I} + {i} \sum_{k} \epsilon_{ijk} \sigma_k,
\end{equation}
\noindent we can immediately see that the inclusion of quadratic terms $L_i L_j$ of the enveloping algebra $U(su(2))$ in the Hamiltonian does not bring anything new to the problem when formulated in terms of $2 \times 2$ matrices. This is because the generators of $su(2)$ correspond to a basis in two-dimensions but this property does not hold anymore for higher dimensions (higher spin representation). It seems that the higher the spin representation the more effect the quadratic terms will have.
In the basis of the spin $1$ representation, with $l=1, m=-1,0,1$ in (\ref{spinbasis}) the expressions in (\ref{quadmetaction}) take the simple form of $3 \times 3$ matrices, so that one is left with 9 equations, complex in general, when imposing the condition (\ref{dysonherm}). Requiring the quadratic metric (\ref{quadmet}) to be invariant under the same anti-linear symmetry we must have $\zeta_\pm = \zeta_1 \pm {i} \zeta_2$, with $\zeta_0, \zeta_1,\zeta_2 ~ \in ~ \mathds{R}$, which makes the metric also Hermitian.

Hermiticity can then be imposed by requiring

\begin{eqnarray}\label{quadconst}
\nonumber
\beta_0 &=& \frac{4 \beta_{++} \zeta _+ \left(8 \zeta _+^2-\sqrt{2-4 \zeta_+^2}+2\right)}{16 \zeta _+^2+1},\\
\nonumber
\beta_+ &=& \frac{\beta_{+0} \left(2 \zeta _+-1\right)}{-2 \zeta _++e^{2 \zeta _0}\left(\zeta _- \left(4 \zeta _+-2\right)+1\right)-1},\\
\nonumber
\alpha_+ &=& -\frac{\alpha_{+0} \left(2 \zeta _++1\right)}{2 \zeta _++e^{2 \zeta _0}\left(\zeta _- \left(4 \zeta _++2\right)+1\right)-1},\\
\zeta _- &=& \frac{\sqrt{2} \zeta _+ \sqrt{1-2 \zeta _+^2}-\zeta _+}{4 \zeta _+^2-1}.
\end{eqnarray}
\noindent with $|\zeta_+| \leq \frac{1}{\sqrt{2}}$. Although the actions (\ref{quadmetaction}) used are representation independent, the Hermiticity constraints above (\ref{quadconst}) were constructed from a $3 \times 3$ representation and therefore are not valid in general. It is interesting to note that no situation for which the quadratic metric coexists with the linear metric could be found. This reinforces the need of more general metric operators as the limitations of the Dyson operator with only linear operators in the exponent becomes clear now. Finally, in Figure 4  we have an example of non-Hermitian Hamiltonian with real eigenvalues for which a quadratic metric, of the type (\ref{quadmet}), can be found.

\begin{figure}[h!]
\begin{center}
\includegraphics[width=0.45\textwidth]{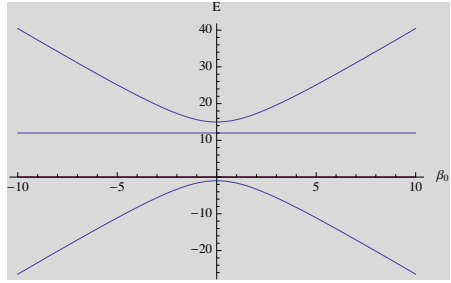}
\fl
\caption{First levels of the real energy spectrum, as a function of $\beta_0$, of the non-Hermitian Hamiltonian $H = {i} \beta_0 L_0 +4\left( 1+ \frac{ {i} \beta_0}{\sqrt{7}-5} \right)
   (L_-^2-L_+^2)+L_0^2+3 (L_- L_+ + L_+L_-)$. This system can be mapped into a Hermitian counterpart with a Dyson operator of the form $\eta = e^{\frac{2-\sqrt{7}}{6}L_-^2}e^{\zeta_{0}L_0^2}e^{\frac{1}{4}L_+^2}$ and has eigenvalues $12$ and  $7 \pm  \frac{\sqrt{81  \beta_0^2+64 \left(431-160 \sqrt{7}\right)}}{16-5\sqrt{7}}$, therefore always real.}
\label{lineargraph}
\end{center}
\end{figure}

\subsection{Swanson diagonalisation}

It is in general not clear why a quadratic non-Hermitian Hamiltonian should allow for a simple $ \eta$ operator of the form (\ref{metansatz}) at all, and we now know that there is a wider class of Hamiltonians for which only a more complicated $ \eta$ operator can be found. In what follows we shall briefly outline how the class of diagonalizable operators can be constructed, and then we will focus on the more general class of operators that have an $ \eta$ operator of the linear form, that does not necessarily diagonalise them.

For the investigation of the quadratic Hamiltonians that can be diagonalised by the $ \eta$ operator with linear exponent, it is convenient to view the action of  $ \eta$ as a generalised Bogoliubov transformation \cite{swanson}. If we make in (\ref{simmatrix}) the identification
\begin{eqnarray}
\nonumber \alpha = 2\lambda_+ \frac{\sinh{\theta}}{\theta} \; ,
\;\;\;\;\;\
\beta = \cosh{\theta}+\lambda_0 \frac{\sinh{\theta}}{\theta} \; , \\
\delta = 2\lambda_- \frac{\sinh{\theta}}{\theta} \; , \;\;\;\;\;\
\gamma = \cosh{\theta}-\lambda_0 \frac{\sinh{\theta}}{\theta} \; ,
\end{eqnarray}
\ni it is also possible to interpret the same transformation as a generalized Bogoliubov transformation \cite{swanson}:
\begin{equation}
\left(
\begin{array}{c}
\tilde{L}_{0} \\
\tilde{L}_{+} \\
\tilde{L}_{-}
\end{array}
\right) = \left(
\begin{array}{ccc}
\gamma \beta +\delta \alpha & -\alpha \beta  & \gamma \delta  \\
-2\delta \beta & \beta^{2} & -\delta^{2} \\
2\gamma \alpha & - \alpha^{2} &  \gamma^{2}
\end{array}
\right) \left(
\begin{array}{c}
L_{0} \\
L_{+} \\
L_{-}
\end{array}
\right),
\end{equation}
\ni with $\beta\gamma-\alpha\delta =1$ to guarantee unity determinant, reducing the number of parameters in the transformation to 3, the same as in the similarity transformation.

For the $su(1,1)$ representation in terms of $K_i$ the interesting feature which allows the diagonalisation of the system \cite{swanson, afkk} is that new creation and annihilation operators. For the representation in terms of $L$ the corresponding new bosons are
%
\begin{equation}
\left(
\begin{array}{c}
d_1 \\
d_2 \\
c_1 \\
c_2
\end{array}
\right) =\left(
\begin{array}{cccc}
\gamma & -\alpha & 0 & 0 \\
-\delta & \beta & 0 & 0 \\
0 & 0 & \beta & \delta \\
0 & 0 & \alpha & \gamma
\end{array}
\right) \left(
\begin{array}{c}
a_1 \\
a_2 \\
a_1^\dag \\
a_2^\dag
\end{array}
\right) ,
\end{equation}

As a result, the transformed operators are written in a very simple form in terms of these new bosonic operators
\begin{equation}
\tilde{L}_0 = \frac{1}{2}(c_1 d_1 - c_2 d_2), \;\;\;\;\;\;
\tilde{L}_+ = c_1 d_2, \;\;\;\;\;\; \tilde{L}_- = c_2 d_1,
\end{equation}

\ni with $[d_1, c_1] =1, [d_2, c_2]=1$ and the remaining
commutators vanishing. Consequently, $\tilde{N}_i = c_i d_i$
behave as number operators and $[\tilde{N}_i, c_i]=c_i,
[\tilde{N}_i, d_i]=-d_i$. If a Hamiltonian can be expressed as
\begin{eqnarray}
\tilde{H} & = & \tilde{a}_0 \tilde{L}_0 + \tilde{a}_{00}
\tilde{L}_0^2 + \tilde{a}_{+-} \tilde{L}_+ \tilde{L}_-  +
\tilde{a}_{-+} \tilde{L}_- \tilde{L}_+ \\
\nonumber & = & \frac{\tilde{a}_{0}}{2}(\tilde{N}_1 - \tilde{N}_2)
+ \frac{\tilde{a}_{00}}{4}(\tilde{N}_1^2 - 2\tilde{N}_1\tilde{N}_2
+ \tilde{N}_2^2) + \\
\nonumber &+&\tilde{a}_{+-}(\tilde{N}_1 +
\tilde{N}_1\tilde{N}_2) + \tilde{a}_{-+}(\tilde{N}_2 +
\tilde{N}_1\tilde{N}_2),
\end{eqnarray}

\ni then, despite being still non-Hermitian ($\tilde{N}_i^\dag
\neq \tilde{N}_i$), it is already diagonalised for some states
$|\tilde{n}_1 \rangle \otimes |\tilde{n}_2 \rangle $ satisfying
$\tilde{N}_i |\tilde{n}_i \rangle = n_i |\tilde{n}_i \rangle$.
Equivalently \cite{ptbec} the states can be represented in
the usual angular momentum basis $| l,m\rangle$ with $l=n_1+n_2$
and $m=n_1-n_2$, such that $\tilde{L}_0
|\tilde{l},\tilde{m}\rangle =
\tilde{m}|\tilde{l},\tilde{m}\rangle$ and $\tilde{L}_\pm
|\tilde{l},\tilde{m}\rangle = \sqrt{(\tilde{l}\mp
\tilde{m})(\tilde{l} \pm \tilde{m}+1)}|\tilde{l},\tilde{m}\pm 1
\rangle$.

The conditions for which the transformation brings the Hamiltonian
into this diagonal form can be obtained from \cite{afkk} having in
mind that $K_0 \rightarrow -L_0,K_+ \rightarrow L_-,
K_- \rightarrow  - L_+ $. For the linear case, for example,
the constraint obtained is almost the same as for the similarity
transformation (\ref{simthetalin}):
\begin{eqnarray}\label{bogthetalin}
\nonumber
\Gamma _0\frac{ \tanh (\theta )}{\theta }=\frac{\alpha _+ \lambda +\beta _+ \Gamma }{\alpha _+ \Gamma -\beta _+ \lambda }
\end{eqnarray}

\ni so that when $\Gamma_0 =0$ one must have $\alpha _+ \lambda +\beta _+ \Gamma=0$, a condition which can be used to fix the ambiguities in the metric. In this case, the metric will also diagonalise the Hamiltonian.

\section{Conclusion and Outlook}

In this paper we carried out the construction of metrics and Hermitian counterparts associated to a certain class of quadratic $su(2)$ non-Hermitian Hamiltonians by using a Lie algebraic formulation. A method first employed for systems with infinite dimensional representation is now applied to problems admitting finite dimensional representations. In this way our methods and results can be successfully checked with explicit diagonalisation. The symmetry present in such systems is not  $\mathcal{PT}$ but still it was possible to construct non-Hermitian models with real spectra by determining the metric operator and isospectral Hermitian counterparts.
In fact, examples of non-$\mathcal{PT}$-symmetric models with real eingenvalues can be found in e.g. \cite{cannatacomplex, andrianovsusy, ddtrealityproof} but the construction of Dyson operators, as done in this manuscript, corresponds to an important step towards a complete understanding of the non-Hermitian structure of the Hamiltonian.

The equivalence of the spectra of the two Hamiltonian partners, the original non-Hermitian and the transformed Hermitian one, was shown in comparative graphs encoding the algebraic structure determined in exact form.
We have taken our metric ansatz beyond the linear operators in the exponent and explored the possibility of having a more complicated ansatz, quadratic in the $su(2)$ generators. The difficulties  imposed by this metric regarding the commutation relations of quadratic terms were overcome with the use of a finite-dimensional representation. We have used studied a $3 \times 3$ example which has the advantage of being simple enough to allow for a solution but at the cost of not necessarily having a generic representation independent result.

Finally, we have emphasised that the anti-linear symmetry of the Hamiltonians investigated in this manuscript correspond just to one of three possibilities presented and there are other different non-Hermitian Hamiltonians expressed as linear and quadratic combinations of $su(2)$ generators which can have real spectra. The other possible realizations of such a symmetry still remains to be investigated in more detail, in particular the appealing situation where \pt-symmetry comes naturally combined with charge conjugation.
Metrics and Hermitian counterparts for non-Hermitian Hamiltonians described by both $su(2)$ and $su(1,1)$ generators may also be treated with
the knowledge of the transformations presented here and in \cite{afkk}, in combinations which are extensions of Jaynes-Cummings models \cite{jcm, kundujcm}.

\appendix

\section{Metrics with $\Gamma_0\neq0$}

\subsection{$\Gamma=0$}

Letting aside the condition that $\eta^\dag = \eta$, we can still
impose some simplifications by vanishing one of the parameters in
the metric. Again we introduce some proportional factor relating
the remaining parameters, $\Gamma_0=\lambda \nu$, such that
$T=2\lambda[ {i}\nu L_0 + L_+ + L_-]$. The solutions in this case
are expressed as
\begin{eqnarray}
Y \equiv \frac{\rm tanh(\theta)}{\sqrt{4-\nu^2}} = \frac{4(\beta_{+0}-2\beta_+)
\pm \Delta}{2((2\alpha_+ - \alpha_{+0})\nu - \beta_0(\nu^2+4))}\\
\nonumber
\Delta = \sqrt{16(\beta_{+0}-2\beta_+)^2+4\beta_0(2\alpha_+ - \alpha_{+0})\nu - \beta_0(\nu^2+4)}
\end{eqnarray}
\ni with $\theta = \lambda\sqrt{4-\nu^2}$ and
\begin{eqnarray}
\fl
\nonumber \alpha_{00} & = & 2\alpha_{+-} + \frac{\alpha_{+0}}{8Y} + \frac{Y
\alpha_{+0}}{2} -
\frac{\beta_{+0}+4Y^2\beta_{+0}-8Y\beta_{++}}{8Y^2 \nu}+ \\
\fl
& -& \frac{(\beta_{+0}-8Y\beta_{++})\nu}{8} + \frac{\alpha_{+0}Y
\nu^2}{8} +
\frac{4Y(\alpha_{+0}-Y\beta_{+0}\nu)}{1+Y^2(4+\nu^2)}\\
\fl
\beta_{++} & = & \frac{ \nu
(\alpha_{+0}(1+Y^2(4+\nu^2))-16Y\alpha_{++})}{8(1-Y^2\nu^2)}
+\frac{\beta_{+0}(1+Y^2(4+\nu^2))}{8Y(1-Y^2\nu^2)}
\end{eqnarray}


There is also another set of solutions for which one cannot take
the $\nu \ra 0$ limit
\begin{eqnarray}
\begin{array}{rcl}
\rm tanh (\theta)&=&\pm \frac{1}{\nu}\sqrt{4-\nu^2} \\
\alpha_+ &=& \frac{\alpha_{+0}}{2} +  \frac{\beta_0(\nu^2+2)}{2\nu}  \mp \left( \beta_0 - \frac{\beta_{+0}}{2} \right)\\
\alpha_{++} &=& \pm \frac{(\alpha_{+0} \pm \beta_{+0})(2+\nu^2)}{8\nu}\\
\beta_{++} &=& \pm \left( \frac{\alpha_{00}}{2}-\alpha_{+-} \right) -
\frac{(\alpha_{+0} \mp \beta_{+0})(4+12\nu^2+\nu^4)}{8\nu(2+\nu^2)}
\end{array}
\end{eqnarray}

\subsection{$\lambda=0$}

For this other two-parameters metric which remains to be analyzed
we use $\Gamma_0=\Gamma \nu$, and the argument in
the metric becomes $T=2 {i} \Gamma[\nu L_0 + L_+ - L_-]$.
\begin{eqnarray}
Y \equiv \frac{\rm tanh(\theta)}{\sqrt{4-\nu^2}} =
\frac{4(\alpha_{+0}-2\alpha_+) \pm \Delta}{2(4(\beta_{+0}-2\beta_+)\nu +\beta_0(\nu^2+4))} \\
\nonumber
\Delta =  \sqrt{16(\alpha_{+0}-2\alpha_+)^2-4\beta_0(4(\beta_{+0}-2\beta_+)\nu +
\beta_0(\nu^2+4))}
\end{eqnarray}

\ni with $\theta = \Gamma\sqrt{4-\nu^2}$ and
\begin{eqnarray}
\fl
\nonumber \alpha_{00}&=&2(\alpha_{+-}-\alpha_{++}) -
\frac{\beta_{+0}}{4Y} + \frac{\alpha_{+0}\nu}{4} +
\frac{(1+2Y^2)(\alpha_{+0}-\beta_{+0})-8Y\alpha_{++}}{4Y(Y\nu-1)}+\\
\fl
&+&\frac{(1+2Y^2)(\alpha_{+0}+\beta_{+0})+8Y\alpha_{++}}{4Y(Y\nu+1)}+
\frac{4Y(\beta_{+0}+Y\alpha_{+0}\nu)}{1+Y^2(4+\nu^2)}\\
\fl
\beta_{++}&=& \frac{ \nu
(16Y\alpha_{++}+\beta_{+0}(4+\nu^2))}{8(Y^2\nu^2-1)}
- \frac{\alpha_{+0}(1+Y^2(4+\nu^2))}{8Y(Y^2\nu^2-1)}
\end{eqnarray}


There is also another set of solutions for which one cannot take
the $\nu \ra 0$ limit
\begin{eqnarray}
\begin{array}{rcl}
\rm tanh (\theta)&=&\pm \frac{1}{\nu}\sqrt{4-\nu^2} \\
\alpha_+ &=& \frac{\alpha_{+0}}{2} \pm \left( \beta_0 - \frac{\beta_{+0}}{2}
- \frac{\beta_0(\nu^2+2)}{2\nu}  \right)\\
\alpha_{++} &=&  \frac{(\alpha_{+0} \mp \beta_{+0})(2+\nu^2)}{8\nu}\\
\beta_{++} &=& \mp \left( \frac{\alpha_{00}}{2}-\alpha_{+-} -
\frac{(\alpha_{+0} \pm \beta_{+0})(4+12\nu^2+\nu^4)}{8\nu(2+\nu^2)} \right)
\end{array}
\end{eqnarray}

\ \\

\section*{Acknowledgments}
The author would like to acknowledge gratitude to A. Fring and E.M. Graefe for initial participation in this project, providing very useful comments and suggestions. I am also grateful for discussions with H. J. Korsch and the hospitality at Technical University Kaiserslautern, where this work was started.
P.E.G.A. was partially supported by a City University London research studentship and EPSRC.

\ \\

\end{document}